\documentclass[amsmath,amssymb,prl,twocolumn,notitlepage]{revtex4-2}

\def\be{\begin{equation}}
\def\ee{\end{equation}}
\usepackage{graphicx}
\usepackage{dcolumn}
\usepackage{bm}
\usepackage{float}
\usepackage{amsmath}
\usepackage{amssymb}
\usepackage{lineno}
\usepackage{braket} 
\usepackage[bookmarks=true,
colorlinks=true,
linkcolor=blue,
urlcolor=blue,
citecolor=blue,
bookmarks=true,
hyperindex=true
]{hyperref}
\usepackage{natbib}

\begin{document}


\title{Macroscopic superposition of vortex states in a matter wave}

\author{Lingran Kong$^{1}$}

\author{Tianyou Gao$^{1}$}
\thanks{Corresponding author: 602gty@sina.com}

\author{Shi-Guo Peng$^{1}$}

\author{Nenghao Dong$^{1,5}$}

\author{Lijie Zhao$^{1,5}$}

\author{Lushuai Cao$^{2,4}$}

\author{Guangshan Peng$^{2}$}

\author{Wenxian Zhang$^{3}$}

\author{Mingsheng Zhan$^{1,4}$}

\author{Kaijun Jiang$^{1,4}$}
\thanks{Corresponding author: kjjiang@wipm.ac.cn}

\affiliation{$^{1}$ State Key Laboratory of Magnetic Resonance and Atomic and Molecular Physics, Innovation Academy for Precision Measurement Science and Technology, Chinese Academy of Sciences, Wuhan 430071, China}

\affiliation{$^{2}$ MOE Key Laboratory of Fundamental Physical Quantities Measurement and Hubei Key Laboratory of Gravitation and Quantum Physics, PGMF and School of Physics, Huazhong University of Science and Technology, Wuhan 430074, China}

\affiliation{$^{3}$ School of Physics, Hangzhou Normal University, Hangzhou, Zhejiang 311121, China}

\affiliation{$^{4}$Wuhan Institute of Quantum Technology, Wuhan 430206, China}

\affiliation{$^{5}$ University of Chinese Academy of Sciences, Beijing 100049, China}

\date{\today}

\begin{abstract}	
Generating the vortex-state superposition in a matter wave is demanded in many quantum processes such as quantum memory and quantum metrology.
Here we report the experimental generation of macroscopic superposition of vortex states in ultracold quantum gases. By transferring an optical vortex-state superposition to the center-of-mass rotational state of ultracold atoms using the Raman coupling technique, we realize two-vortex and three-vortex superposition states in quantum gases, demonstrating the high dimensionality of the vortex state. We show the controllability of the superposition states on the Bloch sphere. The lifetime of the vortex superposition state in quantum gases is as large as 25 ms, about two orders of magnitude longer than the storage time in atomic ensembles. 
This work 
paves the way for high dimensional quantum processing in matter waves. 

\end{abstract}
	
\maketitle
		
In quantum information processing, the information capacity of a quantum system is given by $\log_{2}{d}$ and the cloning fidelity (the overlap of the clones and the original state) is limited to the upper bound $F_{\textrm{clon}}^{d} = \frac{1}{2}+\frac{1}{1+d}$, where $d$ is the number of dimensions \cite{Zeilinger2019LightOpticalVortices, Yuan2019LightOpticalVortices, Dennis2021NaturePhotonVortex}. Compared to the general freedom with two orthogonal components (for example the polarization of photons), the orbital angular momentum (OAM) with discrete integer numbers offers a higher dimensional Hilbert space. So the vortex state which carries OAM could help us to obtain a high information capacity and enhance the robustness against eavesdropping and quantum cloning. Coherent superposition state is the basic unit of the quantum information. Vortex-state superpositions have been realized in photons \cite{Zeilinger2001NatureOAM, Zeilinger2012ScienceOAM, Zeilinger2017PRLOAM},
forming the flying qudits. However, generating the vortex-state superposition in a matter wave has so far not been realized, which is demanded in many quantum processes such as quantum memory and quantum metrology \cite{Ahufinger2018NJPQuantumSensor, Dowling2012JMOMatterWaveGyroscopy, PRA2016DowlingSagnacInterferlometer, PRResearch2020HornbergerAngularInterference, Haine2016PRLinterferometer, Haine2016PRArotation}. In this regard, dilute atomic Bose-Einstein condensates (BEC), which are macroscopic coherent objects cooled to a nearly zero temperature, become ideal candidates.

Vortices have been produced soon after the experimental realization of the atomic BEC \cite{Ketterle1999PRLCriticalVelocity, Cornell1999PRLVortice}. Different techniques including magnetic scanning \cite{Ketterle2003PRLCorlessVortex, Shin2012PRLBECSkymion}, atomic rotating \cite{Ketterle2007PRLSuperfluidRotating, Cornell2001PRLBECRotating} and trap stirring \cite{Ketterle2001PRLBECStirring, Dalibard2000PRLBECStirring} have been applied to excite vortices. Exotic quantum states like Dirac monopole and vortex lattice emerged with the association of vortices \cite{Ketterle2001ScienceVortexLattice, Ketterle2005NatureFermiSuperfluid, Ketterle2006ScienceImbalancedFermiSuperfluid, Hall2014NatureDiracMonopole, Hall2015ScienceIsonlatedMonopole}. 
However, directly exciting a coherent superposition of vortex states in ultracold atoms is very challenging. 
A superposition of vortex states in optical beams has been well produced and controlled \cite{Zeilinger2019LightOpticalVortices, Yuan2019LightOpticalVortices}, which could become an available tool to manipulate ultracold atoms. The single-vortex state of a Laguerre-Gaussian (LG) beam has been transferred to quantum gases through the Raman process \cite{Phillips2006PRLRotation, Bigelow2009PRLBECSkyrmion, Hadzibatic2012PRAAQantizedSupercirrent, Jiang2019PRLSOAMC, Jiang2022npjQuantumInterferometer, Bigelow2009PRLBECVortex, Lin2018PRLSOAMC, Lin2018PRLRotaingBEC}. Particularly, J. Dowling and his colleagues have proposed to realize a coherent superposition of vortex states in quantum gases via an optical angular momentum beam \cite{Dowling2005PRLvortexQubit, Dowling2005PRAvortexQubit}.

Here we experimentally generate macroscopic superposition of vortex states in quantum gases by transferring an optical vortex-state superposition to the center-of-mass rotational state of ultracold atoms. The two-vortex and three-vortex superpositions are generated, respectively, demonstrating the high dimensionality of the vortex state. The three key parameters of the superposition state, namely winding number, relative amplitude and relative phase, are demonstrated to be all controllable on the Bloch sphere. The lifetime of the vortex superposition state in external rotational states of quantum gases is about two orders of magnitude larger than the storage time in internal spin states of atomic ensembles \cite{Guo2013NCSinglePhotonMeory, Guo2014PRAMemoryOAM, Laurat2013NaturePhononOAMmemory, Guo2022PRLLongMemoryOAM}. 
This is the first time to realize and manipulate the vortex-state superposition in a matter wave.

\begin{figure}[htbp]
\includegraphics[width=8cm]{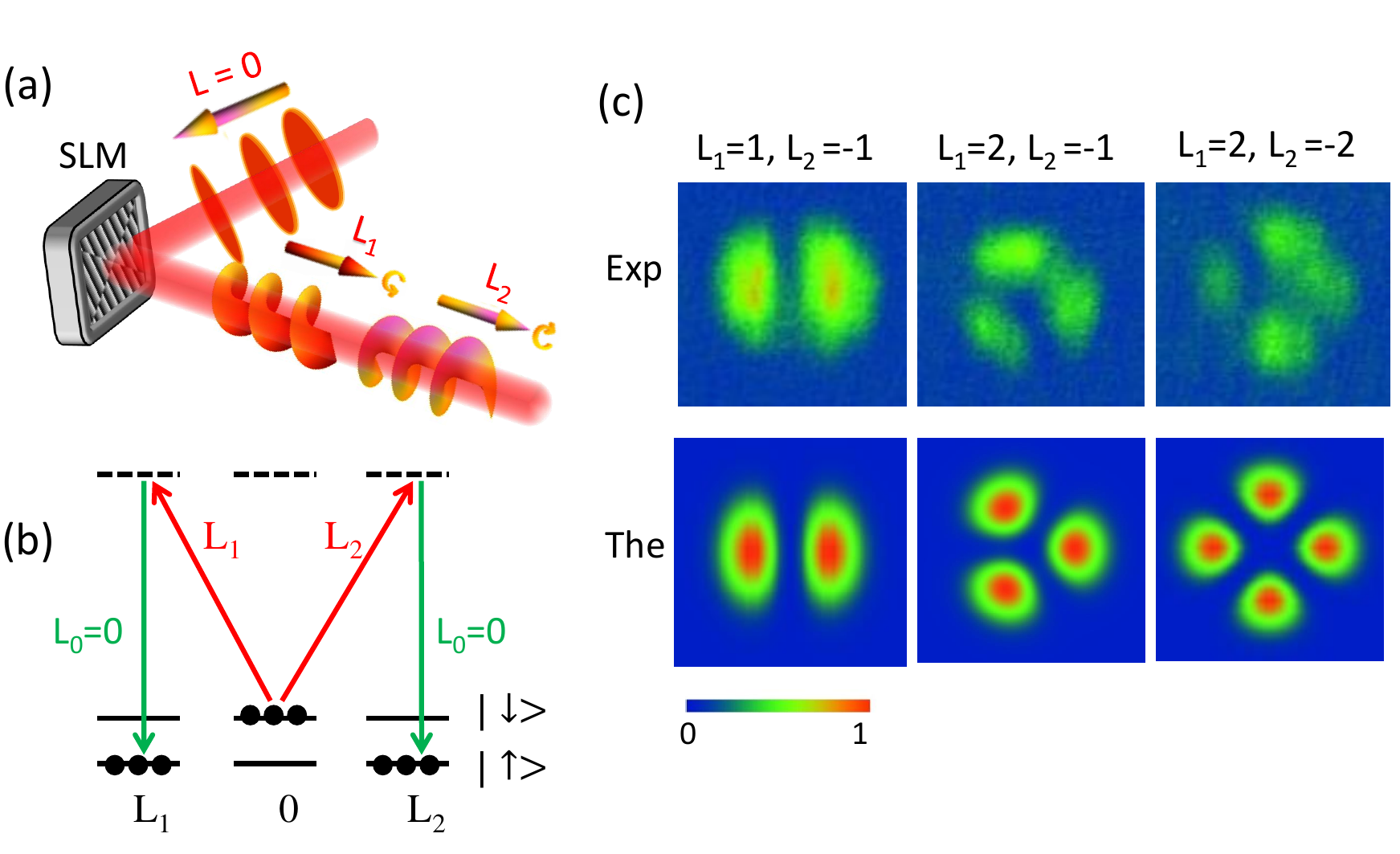}
\caption{The scheme to generate a vortex superposition in ultracold atoms. (a) 
A SLM converts a Gaussian laser beam $L=0$ to a LG beam carrying a superposition of two vortex states $\left|L_{1}\right>$ and $\left|L_{2}\right>$. The optical superposition state is controlled by a computer-generated hologram input to the SLM. 
(b) Atomic energy levels with interactions of laser beams. The Raman beams are composed of a LG beam with a superposition of two vortex states ($\left|L_{1}\right>$ and $\left|L_{2}\right>$) and a Gaussian beam $L_{0}=0$. Atoms initially populated in the state $|0, \left\downarrow\right>$ are transferred to a two-vortex superposition state composed of $|L_{1}, \left\uparrow\right>$ and $|L_{2}, \left\uparrow\right>$ through the two-photon Raman process. 
(c) 
Three typical interference patterns of atoms in the spin state $\left|\uparrow\right>$. 
The winding numbers of the vortex superposition states are $\left(L_{1}=1, L_{2}=-1\right)$, $\left(L_{1}=2, L_{2}=-1\right)$ and $\left(L_{1}=2, L_{2}=-2\right)$, respectively. The first row is for the atomic images in the experiment and the second row is for the corresponding theoretical calculations \cite{Supplemental_superposition}. 
The color bar indicates the normalized optical density (OD) for atoms.}
\label{Fig1}	
\end{figure}

We prepare a spherical $^{87}\textrm{Rb}$ BEC in an optical dipole trap \cite{Jiang2019CPBSphericalBEC, Jiang2019PRLSOAMC, Jiang2022npjQuantumInterferometer}. 
The amplitude and sign of the winding number of an optical vortex state are determined by using the angular interference technique \cite{Jiang2022OCVortexRotation}. We shine an optical Raman pulse onto atoms after a time of flight (TOF) of 3 ms. 
As shown in Fig. \ref{Fig1}, 
the Raman beams are composed of a LG beam and a Gaussian beam $L_{0}=0$. As previously proposed \cite{Dowling2005PRLvortexQubit, Dowling2005PRAvortexQubit}, the LG beam carries a two-vortex superposition $\left|L_{1}\right>+\alpha e^{i\phi}\left|L_{2}\right>$ where $L_{1, 2}$ are the winding numbers, $\phi$ is the relative phase and $\alpha$ is the relative amplitude. Through the Raman process where atoms initially in the spin state $\left|\downarrow\right>$ flip to the spin state $\left|\uparrow\right>$ (see Fig. \ref{Fig1}(b)), the vortex-state superposition is transferred to the center-of-mass rotational state of quantum gases. The quantum state of the BEC is composed of vortex and spin states, and could be represented as  $\{|L,\sigma\rangle\}$ with $L\in\{0,\pm 1,\pm 2,...\}$ and $\sigma \in \{\uparrow,\downarrow\}$, where $|\left\uparrow\right\rangle=|5S_{1/2},F=1,m_F=0\rangle$ and $|\left\downarrow\right\rangle=|5S_{1/2},F=1,m_F=-1\rangle$ are two ground internal states of atoms. The manifolds of internal states $5P_{1/2}$ and  $5P_{3/2}$ are the intermediate states of the two-photon Raman transitions. For a short period of the Raman pulse, i.e., $t \leq50\mu \textrm{s}$, the Hamiltonian could be truncated to the Hilbert space composed of only three states $\{|L=0,\downarrow\rangle,|L=L_1,\uparrow\rangle,|L=L_2\uparrow\rangle\}$. 
The projection of the wavefunction on the spin state $\left|\uparrow\right>$ is written as (see calculation details in Supplemental Material \cite{Supplemental_superposition}) 
 \begin{equation}\label{eq:OAMsuperposition}
\langle \uparrow | \Psi (t) \rangle = A(t) (|L = L_1 \rangle + \alpha \frac{\Omega_{L_2}}{\Omega_{L_1}} e^{i \phi} |L = L_2 \rangle)
\end{equation}

\noindent where $A(t)$ denotes the amplitude in the spin state $\left|\uparrow\right>$, $\Omega_{L_{1(2)}}$ refers to the two-photon transition Rabi frequency induced by the Gaussian beam and $L_{1(2)}$ component of the LG beam. If $L_{1}=-L_{2}$, the vortex-state superposition in the spin state $\left|\uparrow\right>$ of atoms is exactly the same as that in the LG beam with $\Omega_{L_{1}}=\Omega_{L_{2}}$.  As shown in Fig. \ref{Fig1}(a), the three parameters ($L_{1, 2}$, $\phi$ and $\alpha$) of the optical vortex-state superposition could be precisely controlled by designing the computer-generated hologram on the spatial light modulator (SLM) \cite{Laurat2013NaturePhononOAMmemory, Lin2013PRAOpticalSuperposition, Guo2014PRAMemoryOAM, Guo2022PRLLongMemoryOAM}. 
In this way, the vortex-state superposition in ultracold atoms is also controllable, which will be proven in this work.

The atoms in spin state $\left|\uparrow\right>$ are imaged with the help of a Stern-Gerlach magnetic field and 
after a TOF of 22 ms. Fig. \ref{Fig1}(c) shows vortex superposition states with controllable winding numbers. Here we set $\phi=0$ and $\alpha=1$. Three optical vortex-state superpositions of $\left|L_{1}=1\right>+\left|L_{2}=-1\right>$, $\left|L_{1}=2\right>+\left|L_{2}=-1\right>$ and $\left|L_{1}=2\right>+\left|L_{2}=-2\right>$ are used to manipulate ultracold atoms, respectively. The number of the interference lobes is determined by the winding number difference $\Delta L = |L_{1}-L_{2}|$. 
The winding numbers $L_{1, 2}$ of the superposition states in atoms are the same as those in optical beam, which is also confirmed by the numerical calculations using the experimental parameters. Using the same method, we have generated vortex superposition states in atoms with other kinds of winding numbers. 
		
\begin{figure}[htbp]
\includegraphics[width=8cm]{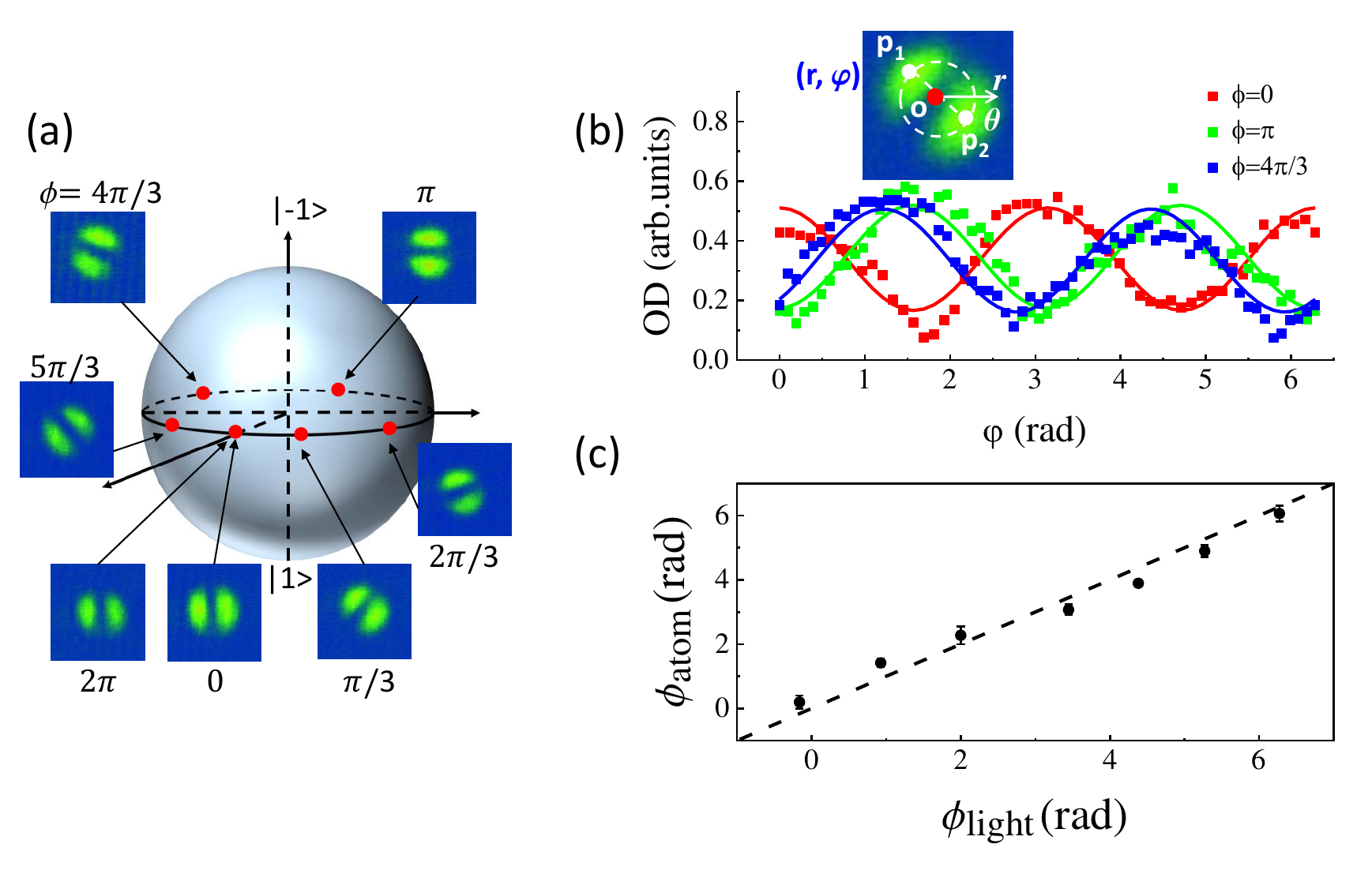}
\caption{Vortex-state superpositions with controllable relative phases. (a) Interference patterns of atoms with the superposition states $|L_1=1 \rangle +  e^{i \phi} | L_2=-1 \rangle$ are schematically displayed on the equator of the Bloch sphere. The relative phase $\phi$ varies with a step of $\pi/3$. (b) Angular interference fringes of ultracold atoms for $\phi=0, \pi, 4\pi/3$, respectively. The solid curves are the consine fitting of the  experimental data, extracting the azimuthal angle of the bright interference lobe $\theta_{\textrm{atom}} = -0.10, -1.54, -1.95$, respectively. Then the relative phase $\phi_{atom}$ can be obtained from Eq. \eqref{eq:relativephase}. The inset schematically shows an interference pattern 
where the inference fringe is along the white dashed circle determined by the two OD maxima $\textrm{p}_{1}$ and $\textrm{p}_{2}$. (c) The relative phase $\phi_{\textrm{atom}}$ in atoms versus $\phi_{\textrm{light}}$ in light. 
The error bar is the standard deviation of 6 to 9 measurements. 
The dashed line represents the calculation of $\phi_{\textrm{atom}} = \phi_{\textrm{light}}$. }
\label{Fig2}
\end{figure}

Figure \ref{Fig2} shows vortex-state superpositions with controllable relative phases. Here we set $L_{1}=-L_{2}=1$ and $\alpha=1$. The interference patterns of atoms are displayed on the equator of the Bloch sphere, as shown in Fig. \ref{Fig2}(a), and rotate clockwise with increasing the relative phase $\phi$. In the cylindrical coordinates $(r, \varphi, z)$ with the radius $r=\sqrt{x^{2}+y^{2}}$ and the azimuthal angle $\varphi$, ultracold atoms are confined in the plane of $z=0$.  
We use a cosine function $\textrm{OD}=\textrm{OD}_{0}+A\cos\left[\Delta L\left(\varphi-\theta\right)\right]$ to fit the angular interference fringe, where $\Delta L = 2$ is the winding number difference, $\textrm{OD}$ is the optical density with a bias $\textrm{OD}_{0}$, and $A$ is the amplitude. $\theta$ is defined as the azimuthal angle of the bright interference lobe in the range $\theta \in \left(-\pi,0 \right]$. 
Figure \ref{Fig2}(b) shows three typical interference fringes for $\phi=0, \pi, 4\pi/3$, where we obtain $\theta_{\textrm{atom}} = -0.10, -1.54, -1.95$, respectively. Then using the relation between the azimuthal angle and relative phase \cite{Jiang2022npjQuantumInterferometer}
\begin{equation} \label{eq:relativephase}
\phi=-\Delta L \theta ,
\end{equation}
we get the relative phase $\phi_{\textrm{atom}}$ in atoms. The relative phase $\phi_{\textrm{light}}$ in light could also be obtained using the same method. We measure the interference fringes for different relative phases in Fig. \ref{Fig2}(c), and it is demonstrated that $\phi_{\textrm{atom}} \approx \phi_{\textrm{light}}$. 

\begin{figure}[htbp]
\includegraphics[width=8cm]{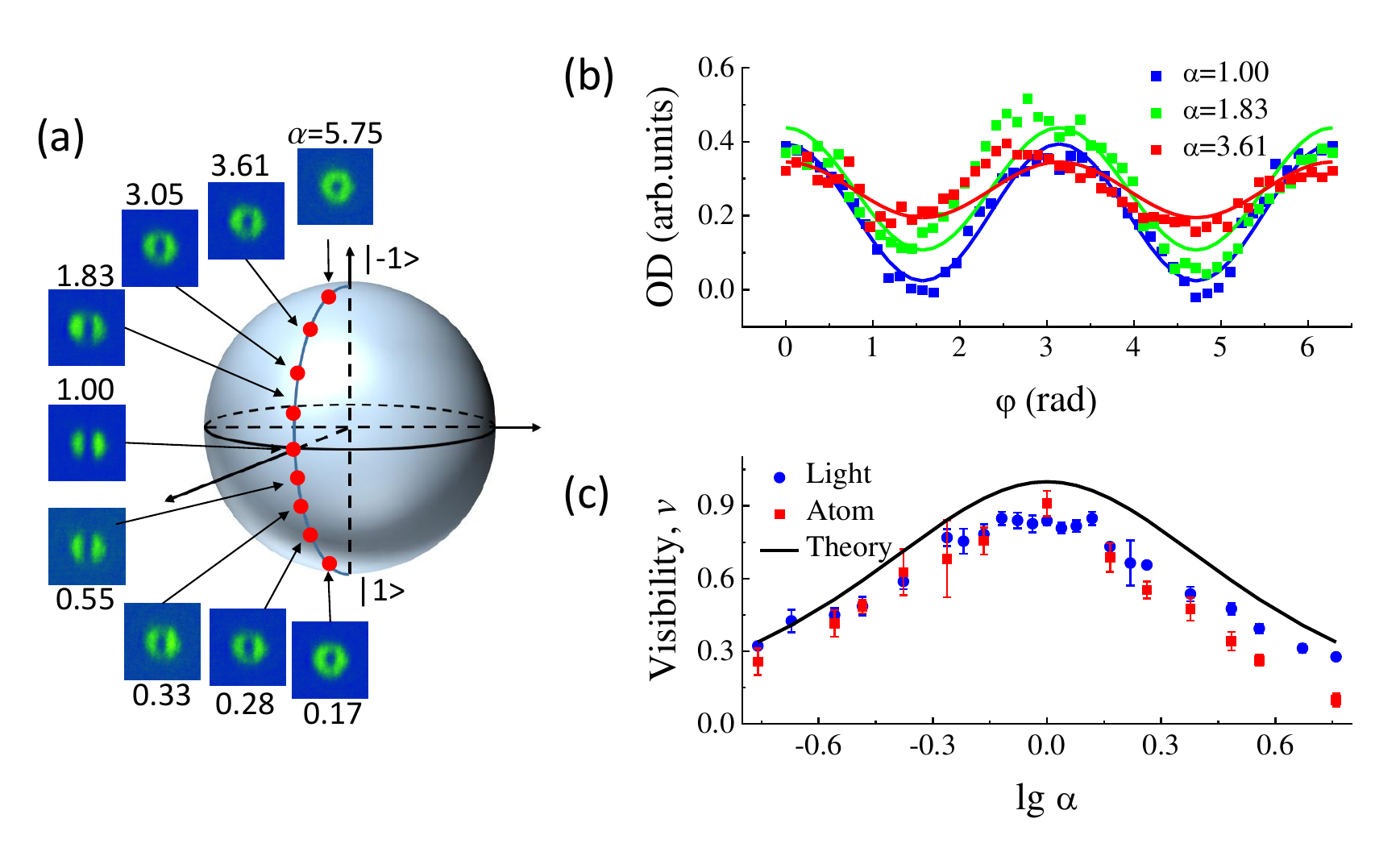}
\caption{Vortex-state superpositions with controllable relative amplitudes. (a) Interference patterns of atoms with the superposition states $|L_1=1 \rangle +  \alpha | L_2=-1 \rangle$ are schematically displayed on the meridian of the Bloch sphere. The relative amplitude $\alpha$ varies in the range $[0.17, 5.75]$. (b) Angular interference fringes of atoms for $\alpha =$ 1.00, 1.83 and 3.61. 
The solid curves are the cosine fitting of the experimental data, extracting the interference visibility $v = $ 0.91, 0.55 and 0.26, respectively. (c) The interference visibility $v$ as a function of the relative amplitude $\alpha$. Note that the parameter along the horizontal axis is $\lg \alpha$. The measurements include atoms (red squares) and light (blue circles). The solid curve is the calculation of Eq. \eqref{eq:visibility}. The error bar is the standard deviation of 6 measurements.  }
\label{Fig3}
\end{figure}

Figure \ref{Fig3} shows vortex-state superpositions with controllable relative amplitudes. Here we set $L_{1}=-L_{2}=1$ and $\phi=0$. The interference visibility $v$ depends on the relative amplitude $\alpha$ as \cite{Dowling2005PRLvortexQubit, Dowling2005PRAvortexQubit}
\begin{equation} \label{eq:visibility}
v=\frac{2\alpha}{1+\alpha^{2}}.
\end{equation}
The interference patterns of atoms are displayed on the meridian of the Bloch sphere, as shown in Fig. \ref{Fig3}(a). When $\alpha \rightarrow \infty$ or 0, the superposition state approaches a single-vortex state $\left|L_{1}=1\right>$ or $\left|L_{2}=-1\right>$, both with a nearly zero interference visibility $v \rightarrow 0$. When $\alpha = 1$,  indicating an equal superposition of two vortex states, the interference visibility becomes maximum $v = 1$. In Fig. \ref{Fig3}(b) we fit the angular interference fringe using a cosine function. 
The interference visibility is calculated as $v\equiv (\textrm{OD}_{\textrm{max}}-\textrm{OD}_{\textrm{min}})/(\textrm{OD}_{\textrm{max}}+\textrm{OD}_{\textrm{min}})$, where $\textrm{OD}_{\textrm{max}}$ and $\textrm{OD}_{\textrm{min}}$ are the maximum and minimum of OD, respectively. For $\alpha =$ 1, 1.83 and 3.61, we obtain $v =$ 0.91, 0.55 and 0.26, respectively.

Figure \ref{Fig3}(c) shows the measured $v$ versus $\alpha$. According to Eq. \eqref{eq:visibility}, $v$ remains the same when $\alpha$ becomes its reciprocal, i.e., $v(\alpha)=v(1/\alpha)$, and then the distribution of $v$ is symmetric about $\lg \alpha =0$. We plot $v$ as a function of $\lg \alpha$. Both interference visibilities in atoms and light ($v_{atom}$ and $v_{light}$) are measured, demonstrating that $v_{atom} \approx v_{light}$. 
The measurements show that $v$ is maximum when $\alpha=1$ and becomes small when any vortex component dominates, having the same variation trend as the calculation of Eq. \eqref{eq:visibility}. Due to the residual background in absorption images, the measured values are some smaller than theoretical calculations \cite{Guo2022PRLLongMemoryOAM}.

\begin{figure}[hbtp]
\includegraphics[width=8cm]{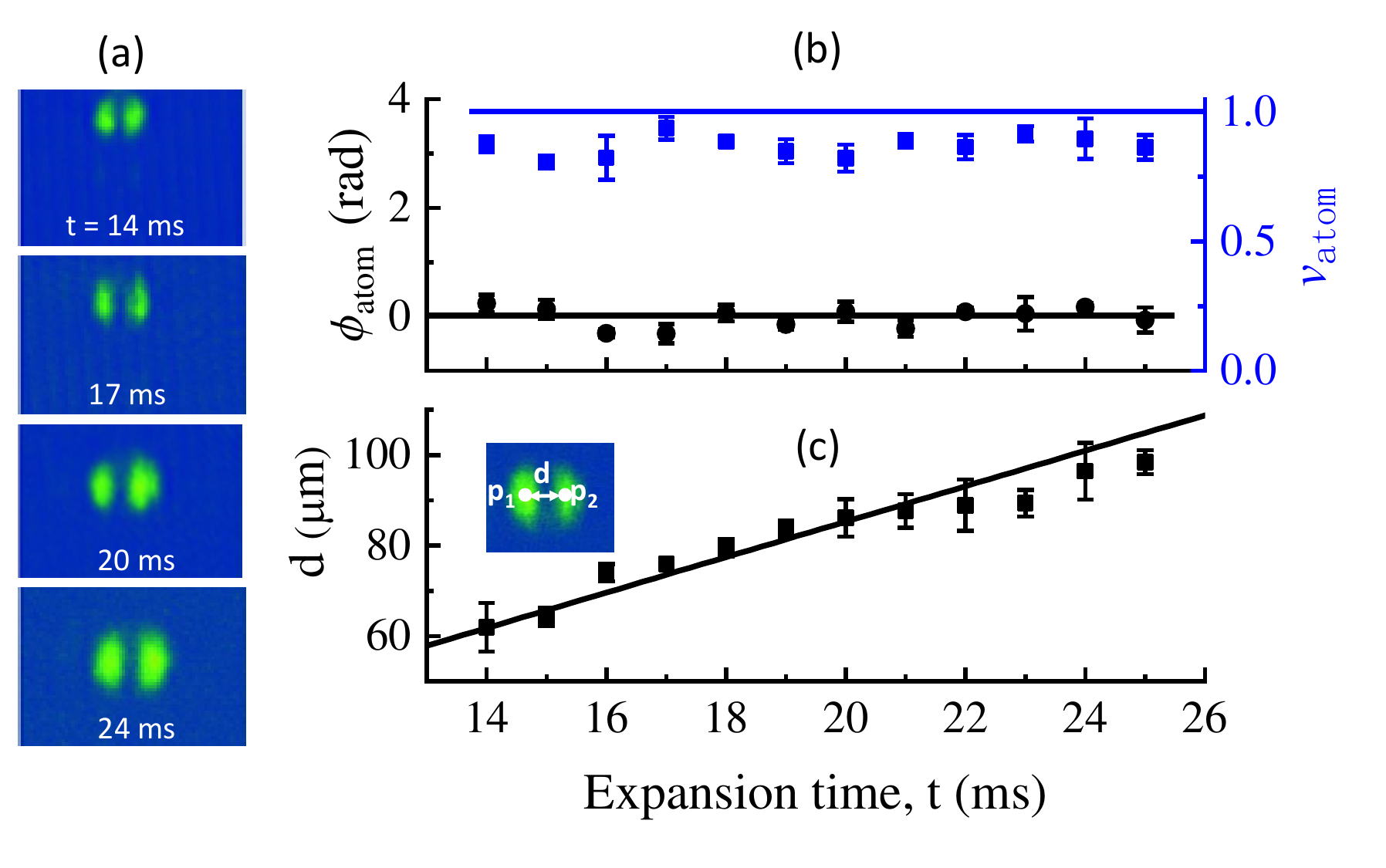}
\caption{Lifetime of the vortex superposition state. (a) Interference patterns of atoms with the superposition state $|L_1=1 \rangle +  \alpha | L_2=-1 \rangle$ during the free expansion. The expansion time $t$ is 14, 17, 20 and 24 ms, respectively. (b) The relative phase $\phi_{\textrm{atom}}$ (black circle) and interference visibility $v_{atom}$ (blue squares) in atoms during the free expansion up to 25 ms. The black solid line is for $\phi_{\textrm{atom}}=0$ and the blue solid line is for $v_{atom} = 1$. (c) Distance $d$ between two interference lobes as a function of the expansion time. 
The solid line represents the theoretical calculation. The error bars are the standard deviations of 6 measurements.} \label{Fig4}
\end{figure}

In Fig. \ref{Fig4} we measure the lifetime of the vortex superposition state in quantum gases. We set $L_{1}=-L_{2}=1$, $\phi=0$ and $\alpha=1$. After the vortex superposition state is prepared in the spin state $\left|\uparrow\right>$, the interference patterns of atoms are probed for different expansion times up to 25 ms. The interference structure with two lobes, as shown in Fig. \ref{Fig4}(a), remains the same during the expansion, indicating the constant winding numbers of $L_{1}=-L_{2}=1$. In Fig. \ref{Fig4}(b), the relative phase remains constant $\phi_{atom} \approx 0$, and the interference visibility $v_{atom} \approx 1$ is unchanged with a constant relative amplitude $\alpha_{\textrm{atom}}\approx 1$. In a word, the three parameters of the vortex-state superposition keeping constant during the atomic expansion demonstrates a minimum lifetime of 25 ms. The expansion time in the experiment is limited by the finite size of the vacuum chamber. In Fig. \ref{Fig4}(c), the distance between two interference lobes increases during the expansion, consistent with the theoretical prediction. Without vortex state, the two lobes should merge together.

The vortex superposition state has been stored in the internal spin excitations of atomic ensembles \cite{Guo2013NCSinglePhotonMeory, Guo2014PRAMemoryOAM, Laurat2013NaturePhononOAMmemory, Guo2022PRLLongMemoryOAM, Guo2023PRLHighdimensionStroage}, and the storage time is on the scale of microsecond, which is limited by dephasing between the two spin states due to different Zeeman shifts caused by the residual magnetic field. The energy difference between two spin states is sensitive to an external magnetic or optical field. Different from the atomic ensemble, the superposition state realized in our experiment is composed of two external center-of-mass rotational states, and exists only in one internal spin state. Consequently the external magnetic or optical field would not change the relative phase between two vortex states, strongly suppressing the dephasing effect. 
We show that the lifetime of the vortex superposition state in quantum gases is about two orders of magnitude longer than the storage time in atomic ensembles.

One advantage of the vortex state is to offer a higher dimensional quantum state (qudit). Here we demonstrate the generation of a three-vortex superposition in quantum gases.
For example, the optical three-vortex superposition is selected to be $\left|\psi \right\rangle= \left| L_{1}= 1 \right> +  e^{i\pi} \left| L_{2}=-1 \right> + \alpha_{3} e^{i\phi_{3}} \left| L_{3}=0 \right>$, where $\alpha_{3}$ and $\phi_{3}$ are the relative amplitude and relative phase of the third state $\left| L_{3}=0 \right>$, respectively. 
For a short period of Raman pulse, the Hamiltonian could be truncated to the Hilbert space composed of only four states $\{|L=0,\downarrow\rangle,|L=0,\uparrow\rangle,|L=\pm 1,\uparrow\rangle \}$. 
The projection of the wavefunction on the spin state $\left|\uparrow\right>$ is written as \cite{Supplemental_superposition} 
\begin{equation} \label{eq:qutrit}
\langle \uparrow | \Psi (t) \rangle = A(t) (|L=1 \rangle + e^{i\pi} |L=-1 \rangle+ \alpha_{3} \frac{ \Omega_{0} }{ \Omega } e^{i \phi_{3}}  |L=0 \rangle )
\end{equation}

\noindent where $A(t)$ denotes the amplitude in the spin state $\left|\uparrow\right>$, $\Omega =\Omega_{L=1} =\Omega_{L=-1}$ refers to the two-photon transition Rabi frequency induced by the Gaussian beam and $L=1(-1)$ component of the LG beam, and $\Omega_{0}$ refers to the Rabi frequency induced by the Gaussian beam and $L=0$ component of the LG beam.

\begin{figure}[hpbt]
\includegraphics[width=8.5cm]{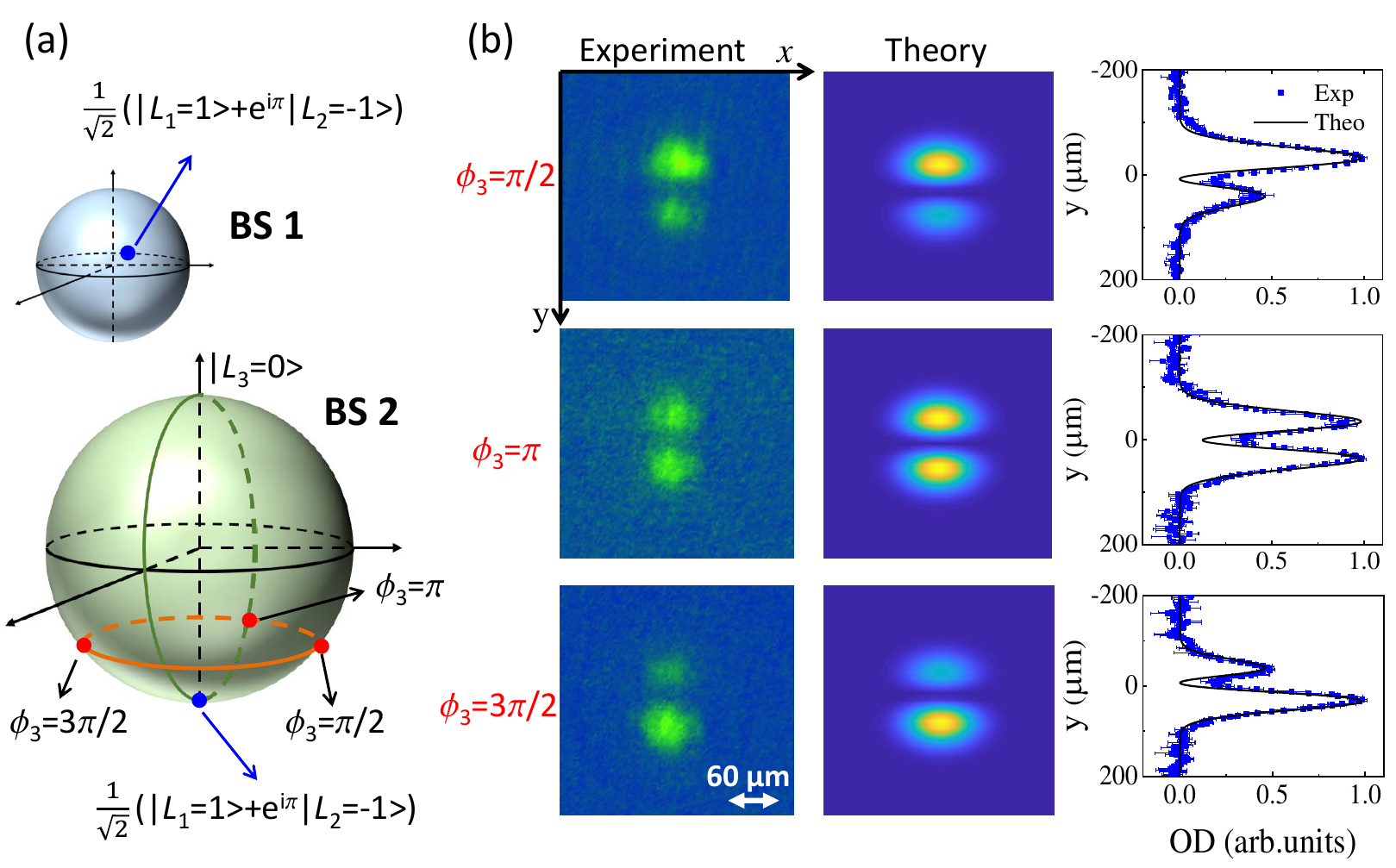}
\caption{Generation of the three-vortex superposition state. (a) The blue circle on the Bloch sphere (BS) 1 denotes the two-vortex superposition state $\left| L_{1}= 1 \right> +  e^{i\pi} \left| L_{2}=-1 \right>$, which is also shown on the south pole of the BS 2 with $\alpha_{3}=0$. The three-vortex superposition with $\alpha_{3}=0.05$ is schematically denoted with the orange ring on the BS 2, and the three red circles represent $\phi_{3} = \pi/2$, $\pi$ and $3\pi/2$, respectively. (b) The interference patterns of atoms carrying the three-vortex superposition states. The left column is for the measurements and the middle column is for the theoretical calculations. The right column shows the normalized integrated OD along $y$ direction, which includes the measurements (blue squares) and theoretical calculations (black solid lines). The error bar is the standard deviation of three measurements. For $\phi_{3} = \pi/2$, the upper interference lobe is brighter than that on the lower half. For $\phi_{3} =3\pi/2$, the situation is opposite. For $\phi_{3} =\pi$, the distribution of the two interference lobes is symmetric.}
\label{Fig5}
\end{figure}
	
To study the effect of 
$\phi_{3}$, we set $\alpha_{3}=0.05$. 
As shown in Fig. \ref{Fig5}(a), the three-vortex superposition state is schematically represented by two Bloch spheres \cite{Supplemental_superposition, Feng_Nature2022_Fourdimension, Feng_elight2024_BlochSphere, Kemp_PRresearch2022_sphere}. For the two-vortex superposition state $\left| L_{1}= 1 \right> +  e^{i\pi} \left| L_{2}=-1 \right>$, two symmetric interference lobes distribute along the vertical direction (see Fig. \ref{Fig2}(a)), and there is a phase singularity line in the center of the atomic cloud,  with a uniform phase of $\pi/2$ on the upper half and a phase of $3\pi/2$ on the lower half \cite{Supplemental_superposition}. Figure \ref{Fig5}(b) explores the variation of the interference pattern when the third component $\left| L_{3}= 0 \right>$ is additionally superposed. 
The symmetry and structure of the interference pattern change with $\phi_{3}$. 
For $\phi_{3} = \pi/2$, constructive interference occurs on the upper half and destructive interference happens on the lower half (see the phase distribution in Supplemental Material \cite{Supplemental_superposition}), and consequently the upper interference lobe is brighter and larger than the lower one. For $\phi_{3} = 3\pi/2$, the situation is exactly opposite. For $\phi_{3} = \pi$, the phase singularity line disappears, and the distribution of the two interference lobes is symmetric on both halves. 
The phase dependence of the interference pattern demonstrates the coherent superposition of three vortex states. We also have generated other three-vortex superposition states with various $\alpha_{3}$ and $\phi_{3}$ (see Supplemental Material \cite{Supplemental_superposition}), 
which shows the controllability on the Bloch sphere.


In conclusion, we have generated a coherent superposition of two and three-vortex states in ultracold quantum gases. 
The superposition state is controllable on the Bloch sphere. The lifetime of the superposition state in quantum gases is about two orders of magnitude longer than the storage time in atomic ensembles. This work paves the way to establish the vortex qudit and vortex entanglement in quantum gases \cite{Dowling2005PRLvortexQubit, Dowling2005PRAvortexQubit}. The vortex superposition state in a matter wave has prospective applications for quantum rotation sensing with a long interrogation time and high robustness against the fluctuation of external fields \cite{Ahufinger2018NJPQuantumSensor, Dowling2012JMOMatterWaveGyroscopy, PRA2016DowlingSagnacInterferlometer, Jiang2022npjQuantumInterferometer}. 

We thank Han Pu for favorite discussions, and Weijia Du for the contribution in the early stage of numerical modeling. This work is supported by the National Key R \&D Program under Grant No. 2022YFA1404102, the National Natural Science Foundation of China under Grants No. U23A2073, No. 12374250, and No. 12121004, Chinese Academy of Sciences under Grant No. YJKYYQ20170025, the Natural Science Foundation of Hubei Province under Grant No. 2021CFA027, and Innovation Program for Quantum Science and Technology under Grant No. 2023ZD0300401.



\newpage
\begin{widetext}

\setcounter{secnumdepth}{3} 

\setcounter{equation}{0}

\setcounter{figure}{0}
	
\renewcommand{\thefigure}{S\arabic{figure}}
\renewcommand{\thetable}{S\arabic{table}}
\renewcommand{\theequation}{S\arabic{equation}}

\section*{Supplemental materials}
	
\section{Experimental Methods} \label{section:I}
\noindent \textbf{Experimental setup.} A $^{87}$Rb BEC is prepared in a spherical optical dipole trap where the wingding number of votex is a good quantum number. The trapping frequency is $\omega=2\pi \times 77.45$ Hz, the asphericity is 3.7$\%$ \cite{Jiang2019CPBSphericalBEC, Jiang2019PRLSOAMC}, the atom number is $N=1.2(1) \times 10^5$, and the temperature is $T \approx 50$ nK. The BEC is initially prepared in the spin state $\left|\downarrow\right>=\left|F=1,m_\textrm{F}=-1\right>$. We shine an optical Raman pulse onto atoms after a time of flight (TOF) of 3 ms, enlarging atomic cloud to increase the coupling strength \cite{Jiang2022npjQuantumInterferometer}. The period of the Raman pulse is about $50\ \mu s$. Two Raman laser beams (one is the Gaussian beam, the other is the Laguerre-Gaussian beam with a superposition of two vortex states) co-propagate across the ultracold atoms. 
The vortex superposition of $\left|L_{1}\right>$ and $\left|L_{2}\right>$ are subsequently present in the final spin state $\left|\uparrow\right>=\left|F=1,m_\textrm{F}=0\right>$. The atoms in $\left|\uparrow\right>$ are imaged after a TOF of 22 ms and with the help of a Stern-Gerlach magnetic field to spatially separate different spin states. A pair of Helmholtz coils produces a bias magnetic field $B_0$, which provides the quantum axis and a large quadratic Zeeman shift $\omega_q=2\pi \times 5.52$ kHz of the Rb ground spin states, making the spin state $|F=1,m_F=+1 \rangle $ far-off resonant with light. 
A probe beam counter-propagates with the Raman beams, detecting the density distribution of the condensate. We use the tune-out wavelength $\lambda=790.02$ nm of the two Raman beams. 
The Radius of the Raman beam at focus is $w_{l}=30$ $\mu$m. The power of each Raman beam is 30 mW.

\noindent \textbf{Controlling the hologram input into the spatial light modulator (SLM).} The Laguerre-Gaussian beam with a vortex or a superposition of vortex states is controlled by a computer-controlled SLM (also see Refs. \cite{Laurat2013NaturePhononOAMmemory, Lin2013PRAOpticalSuperposition, Guo2014PRAMemoryOAM, Guo2022PRLLongMemoryOAM}). To generate a Laguerre-Gaussian beam with a vortex, we generate a data matrix ($1920\ \textrm{columns} \times 1080\ \textrm{rows}$) according to $\exp(il \arctan(y/x))$ using the Matlab software, where $l$ is the winding number of the vortex state and $x$ ($y$) is the coordinate. In addition, we add the periodically changing data on the data matrix, producing a blazed grating to diffract the incident light. Then the data matrix corresponds one-to-one to the pixel of the liquid crystal surface of the SLM. A Gaussian beam is incident on the SLM, and then the diffracted beam becomes a Laguerre-Gaussian beam with a vortex phase term $\exp(i l\theta )$, where $\theta=\arctan(y/x)$ is the azimuthal angle. In a similar way, by setting the data matrix with a vortex superposition phase, we could get a Laguerre-Gaussian beam with a superposition of vortex states.

\noindent \textbf{Determining the local maximum in the interference lobe.} 
Accurately determining the local maximum of each interference lobe is demanded to obtain the angular interference fringe and distance between two interference lobes. The size of a CCD pixel is 6.5 $\times$ 6.5 $\mu \textrm{m}$, and the atomic cloud size after a TOF is about 30 $\times$ 30 pixels. Fluctuation in experimental data can lead to misalignment in determining the maximum points. To smooth the experimental data, we take the mean OD value of adjacent 9 pixels as the value of the midpoint. Using this method, we can determine two maximum points $ \textrm{p}_{1}(x_{1},y_{1})$ and $ \textrm{p}_{2}(x_{2},y_{2})$ in the two lobes, respectively. Then the angular interference path is the circle whose diameter connects $ \textrm{p}_{1}$ and $ \textrm{p}_{2}$, and the distance between $ \textrm{p}_{1}$ and $ \textrm{p}_{2}$ is calculated as
\begin{eqnarray}
	d=\sqrt{\left(x_{1}-x_{2}\right)^2+\left(y_{1}-y_{2}\right)^2}.
\end{eqnarray}

\section{Controllability of the relative phase} \label{section:II}
We generate a two-vortex superposition state $|L_1=1 \rangle +  e^{i \phi} | L_2=-1 \rangle$ where $\phi$ is the relative phase. The winding number difference is $\Delta L = |L_{1}-L_{2}| =2$. The relative phase $\phi_{\textrm{control}}$ controlled by a computer is input to the SLM. The azimuthal angles of the interference lobes in atoms and light ($\theta_{\textrm{atom}}$ and $\theta_{\textrm{light}}$) are both measured, 
as shown in Fig. \ref{figS1}. It is shown that $\phi_{\textrm{control}} \approx -2 \theta_{\textrm{atom}}\approx -2\theta_{\textrm{light}}$. According to the relation $\phi =-\Delta L \theta$, we find that $\phi_{\textrm{atom}} \approx \phi_{\textrm{light}} \approx \phi_{\textrm{control}}$. Data in Fig. \ref{figS1}(a) and (b) are combined to provide the data in Fig. 2(c) of the main text. 

\begin{figure}[hpbt]
\includegraphics[width=0.8\textwidth]{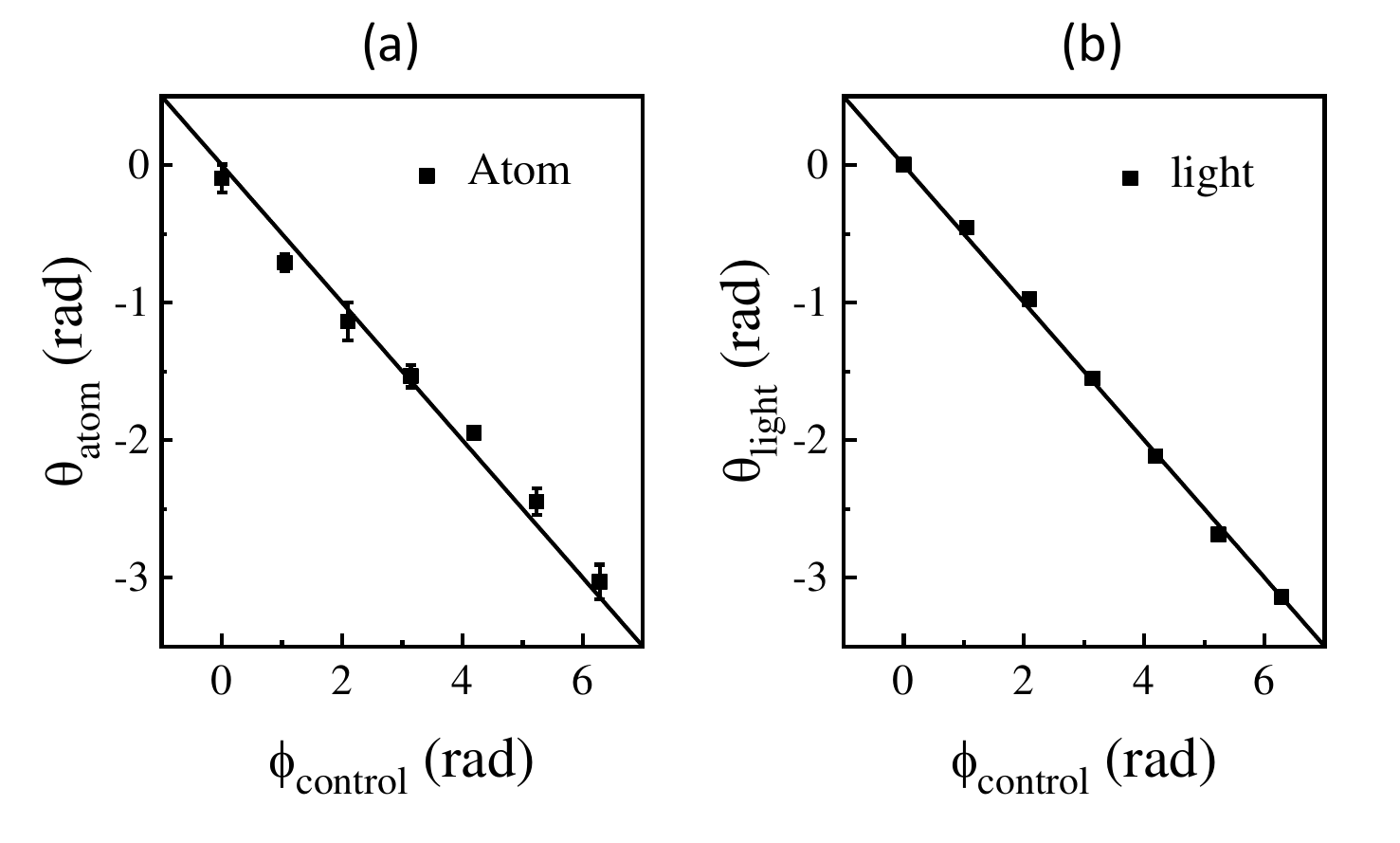}
\caption{
(a) The azimuthal angle of the interference lobe $\theta_{\textrm{atom}}$ in atoms as a function of the relative phase $\phi_{\textrm{control}}$ in control. The black squares are for experiments and the solid line is for the calculation with $\phi_{\textrm{control}} = -2 \theta_{\textrm{atom}}$. The error bars are standard deviations of 6 to 9 measurements. (b) The azimuthal angle of the interference lobe $\theta_{\textrm{light}}$ in light as a function of $\phi_{\textrm{control}}$. The black squares are for experiments and the solid line is for the calculation with $\phi_{\textrm{control}} = -2\theta_{\textrm{light}}$. The error bar for light is smaller than the mark size. }
\label{figS1}
\end{figure}

\section{Phase distribution of atoms with a three-vortex superposition state} \label{section:III}
In Fig. \ref{figS2}. we calculate the phase distribution of atoms with a three-vortex superposition state $\left| L_{1}= 1 \right> +  e^{i\pi} \left| L_{2}=-1 \right> + \alpha_{3} e^{i\phi_{3}} \left| L_{3}=0 \right>$, where $\alpha_{3}$ and $\phi_{3}$ are the relative amplitude and relative phase of the third state $\left| L_{3}=0 \right>$, respectively. For the two-vortex superposition $\left| L_{1}= 1 \right> +  e^{i\pi} \left| L_{2}=-1 \right> $, there is a phase singularity line in the center of the atomic cloud,  with a uniform phase of $\pi/2$ on the upper half and a phase of $3\pi/2$ on the lower half, and the atomic density distribution is symmetric on both halves. To analyze the effect of the third state $\left| L_{3}=0 \right>$, we set a small relative amplitude $\alpha_{3}=0.05$. Because $\alpha_{3}$ is small, the variation of the phase distribution relative to the two-vortex state mainly occurs in the central small area. 
For $\phi_{3}=\pi/2$, a constructive interference occurs on the upper half and a destructive interference happens on the lower half, resulting in the phase singularity line moves downward, and consequently the upper interference lobe is brighter and larger than the lower one; For $\phi_{3} = 3\pi/2$, the situation is opposite; For $\phi_{3} = \pi$, with the phase singularity disappearing, the phase increases continuously from $\pi/2$ to $3\pi/2$ across the central line, and the atomic density distribution is symmetric on both halves.
\begin{figure}[hpbt]
\includegraphics[width=0.8\textwidth]{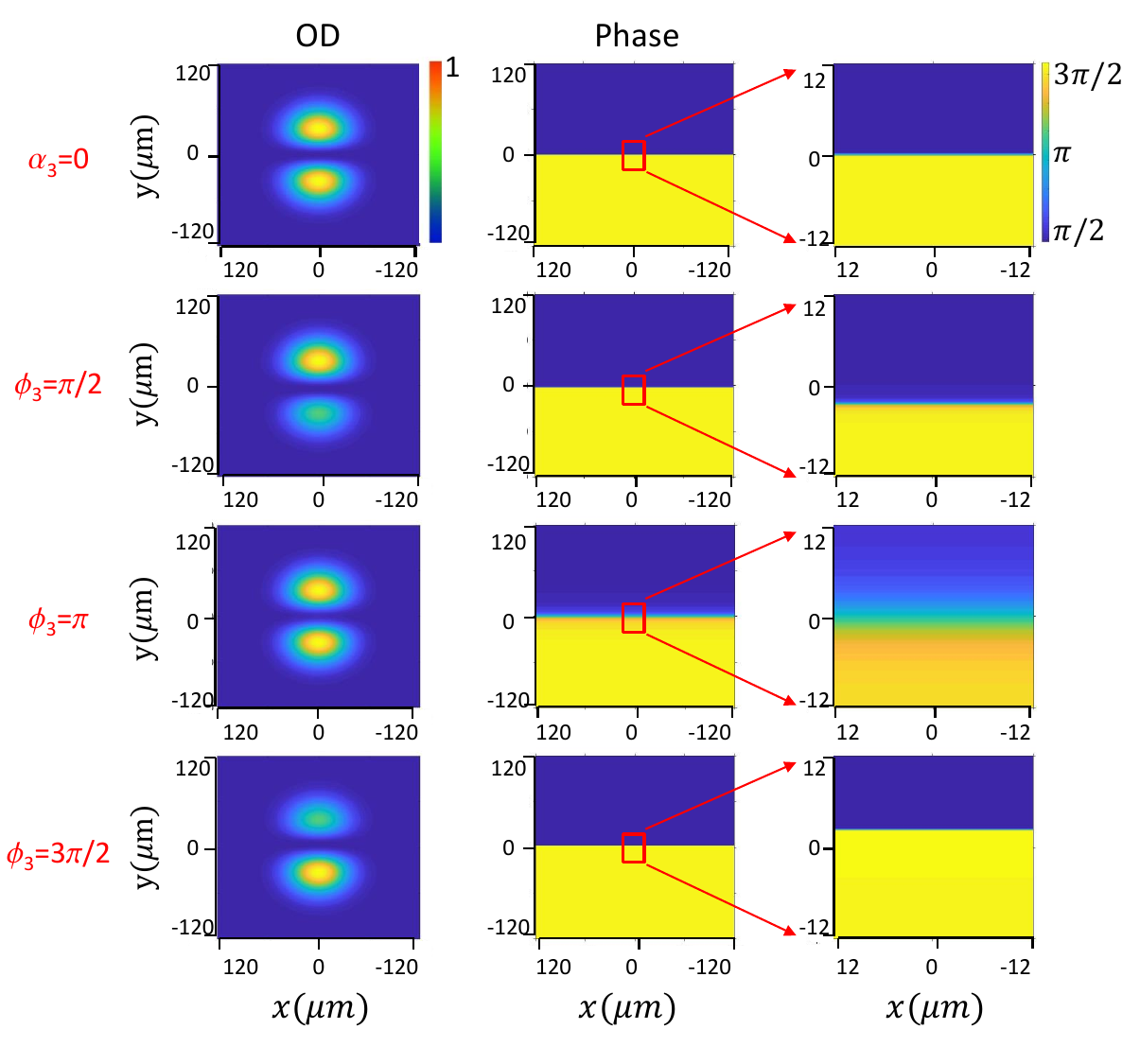}
\caption{Phase distribution of atoms with a three-vortex superposition state $\left| L_{1}= 1 \right> +  e^{i\pi} \left| L_{2}=-1 \right> + \alpha_{3} e^{i\phi_{3}} \left| L_{3}=0 \right>$. The first column is for the optical density (OD) of atoms, the second column is for the phase distribution, and the third column is to zoom the central small area of the second column ($24\ \mu \textrm{m} \times 24\ \mu \textrm{m}$, marked by the red square). For $\alpha_{3}=0$ (the first row), i.e., the two-vortex superposition $\left| L_{1}= 1 \right> +  e^{i\pi} \left| L_{2}=-1 \right> $, there is a phase singularity line in the center of the atomic cloud,  with a uniform phase of $\pi/2$ on the upper half and a phase of $3\pi/2$ on the lower half, and the atomic density distribution is symmetric on both halves. On lower three rows we set $\alpha_{3}=0.05$. For $\phi_{3}=\pi/2$ (the second row), 
the phase singularity line moves downward, and the upper interference lobe is brighter and larger than the lower one; For $\phi_{3} = 3\pi/2$ (the fourth row), the situation is opposite; For $\phi_{3} = \pi$ (the third row), with the phase singularity disappearing, the phase increases continuously from $\pi/2$ to $3\pi/2$ across the central line, and the atomic density distribution is symmetric on both halves.}
\label{figS2}
\end{figure}

\section{Three-vortex superposition states on the Bloch spheres} \label{section:IV}
As shown in Fig. \ref{figS3}, we use two Bloch spheres to represent the three-vortex superposition state. The principle of two Bloch spheres is briefly explained in Sec. \ref{section:VI}. The two-vortex superposition state $\left| L_{1}= 1 \right> + \left| L_{2}=-1 \right>$ is on the Bloch sphere 1, which is also shown on the south pole of the Bloch sphere 2. The superposition of the third state $\left| L_{3}=0 \right>$ is indicated on the Bloch sphere 2. We experimentally generate three-vortex superposition states $\left| L_{1}= 1 \right> + \left| L_{2}=-1 \right> + \alpha_{3} e^{i\phi_{3}} \left| L_{3}=0 \right>$. The density distribution of atoms with the two-vortex superposition state $\left| L_{1}= 1 \right> + \left| L_{2}=-1 \right>$ is symmetric on the left and right halves. After superposing the third state $\left| L_{3}=0 \right>$, the symmetry and structure of the interference pattern strongly depend on $\phi_{3}$ and $\alpha _{3}$. For $\phi_{3}=0$, due to the constructive interference on the right half and destructive interference on the left half, the right interference lobe is bigger than the left one, and the asymmetry becomes smaller when $\alpha _{3}$ decreases. For $\phi_{3} = \pi$, the situation is opposite. For $\phi_{3} = \pi/2$, the atomic distribution on both halves is symmetric regardless of $\alpha _{3}$, and only when $\alpha _{3}$ is small, the two interference lobes are distinguishable. Note that here the relative phase of the second state $\left| L_{2}=-1 \right>$ is $0$, while it is $\pi$ in Fig. 5 of the main text. 
\begin{figure}[hpbt]
\includegraphics[width=0.8\textwidth]{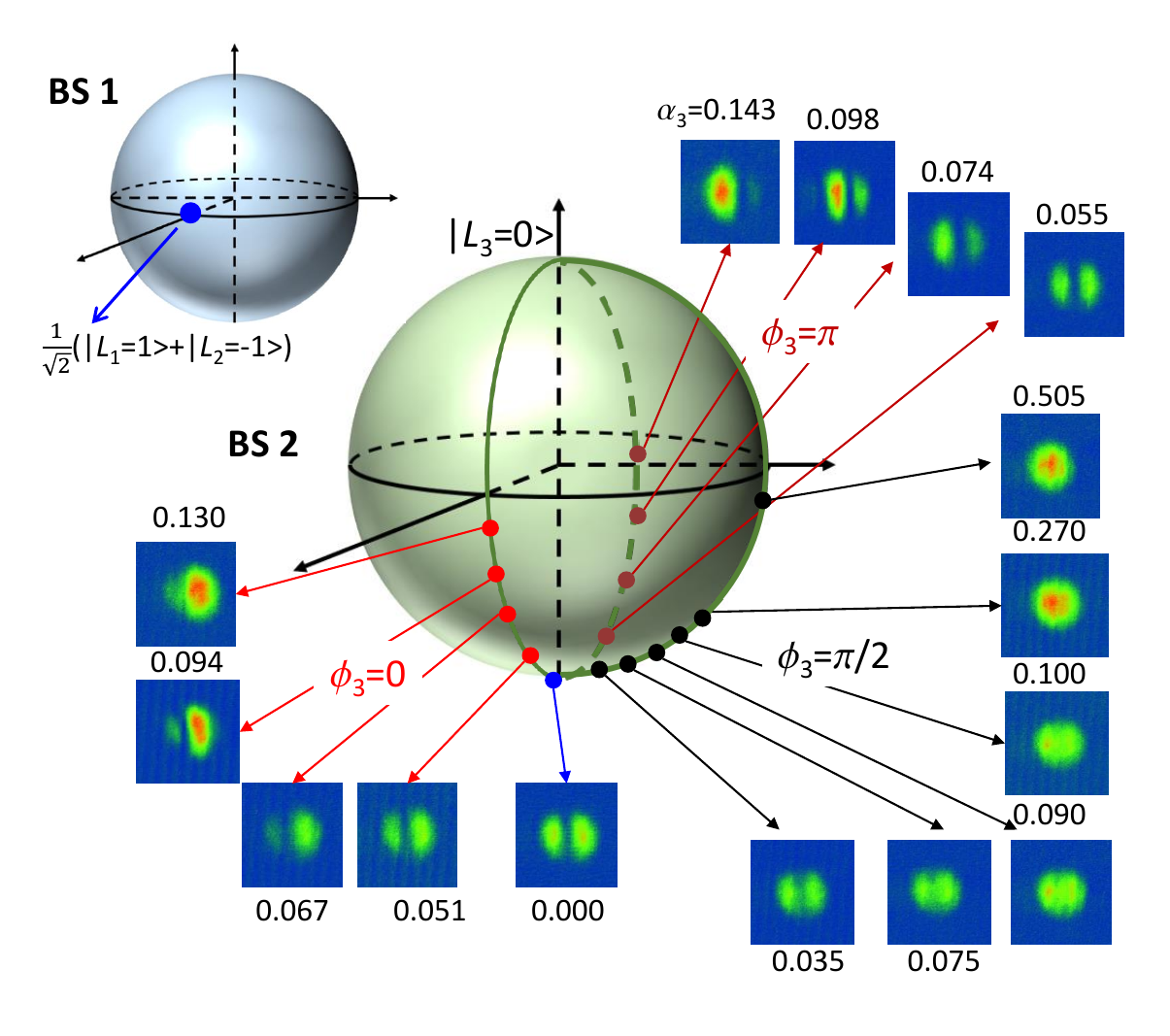}
\caption{Three-vortex superposition with different relative phases and relative amplitudes. The two-vortex superposition state $\left| L_{1}= 1 \right> + \left| L_{2}=-1 \right>$ is denoted with the blue circle on the Bloch sphere (BS) 1, which is also shown on the south pole of the BS 2, i.e., $\alpha_{3}=0$. The there-vortex superposition $\left| L_{1}= 1 \right> + \left| L_{2}=-1 \right> + \alpha _{3} e^{i\phi_{3}} \left| L_{3}=0 \right>$ is represented with two BSs, where $\alpha _{3}$ and $\phi_{3}$ are the relative amplitude and relative phase of the third state $\left| L_{3}=0 \right>$, respectively. Interference patterns of atoms for different $\phi_{3}$ and $\alpha _{3}$ are schematically displayed on the BS 2. The symmetry and structure of the interference pattern strongly depend on $\phi_{3}$ and $\alpha _{3}$. For $\phi_{3}=0$ (red circles), the right interference lobe is bigger than the left one, and the asymmetry becomes smaller when $\alpha _{3}$ decreases. For $\phi_{3} = \pi$ (brown circles), the situation is opposite. For $\phi_{3} = \pi/2$ (black circles), the atomic distribution on both halves keeps symmetric regardless of $\alpha _{3}$, and only when $\alpha _{3}$ is small, the two interference lobes are distinguishable.}
\label{figS3}
\end{figure}

\section{Calculation on the generation of a vortex superposition state in quantum gases} \label{section:V}
\subsection{Generation of a specific two-vortex superposition state with $L=\pm1$} \label{section:IA}
The $^{87}\textrm{Rb}$ atomic Bose-Einstein condensate (BEC) is coupled with a pair of Raman beams, one is the Gaussian and the other is the Laguerre-Gaussian (LG) type. The LG beam carries a vortex superposition state which could be transferred to the BEC during the two-photon Raman process. For convenience, we use $l\hbar$ to denote the orbital angular momentum (OAM) of the LG beam and $L\hbar$ for that of the BEC. During the Raman process with a short period of about $50 \ \mu \textrm{s}$, the BEC could be approximated to be static. We first consider that the LG beam carries a two-vortex superposition state with $l=\pm 1$, as shown in Fig. \ref{sfig1}. The basis states of the BEC could be represented as  $\{|L, \sigma\rangle\}$ with $L\in\{0,\pm 1,\pm 2,...\}$ and $\sigma \in \{\uparrow,\downarrow\}$, where $|\left\uparrow\right\rangle=|5S_{1/2},F=1,m_F=0\rangle$ and $|\left\downarrow\right\rangle=|5S_{1/2},F=1,m_F=-1\rangle$ are two ground internal states of atoms. The manifolds of internal states $5P_{1/2}$ and  $5P_{3/2}$ are the intermediate states of the two-photon Raman transitions. The Gaussian beam has a circular polarization ($\sigma^-$) and the LG beam has a linear polarization ($\pi$).
	
\begin{figure}[htbp]
\centering
\includegraphics[width=0.8\textwidth]{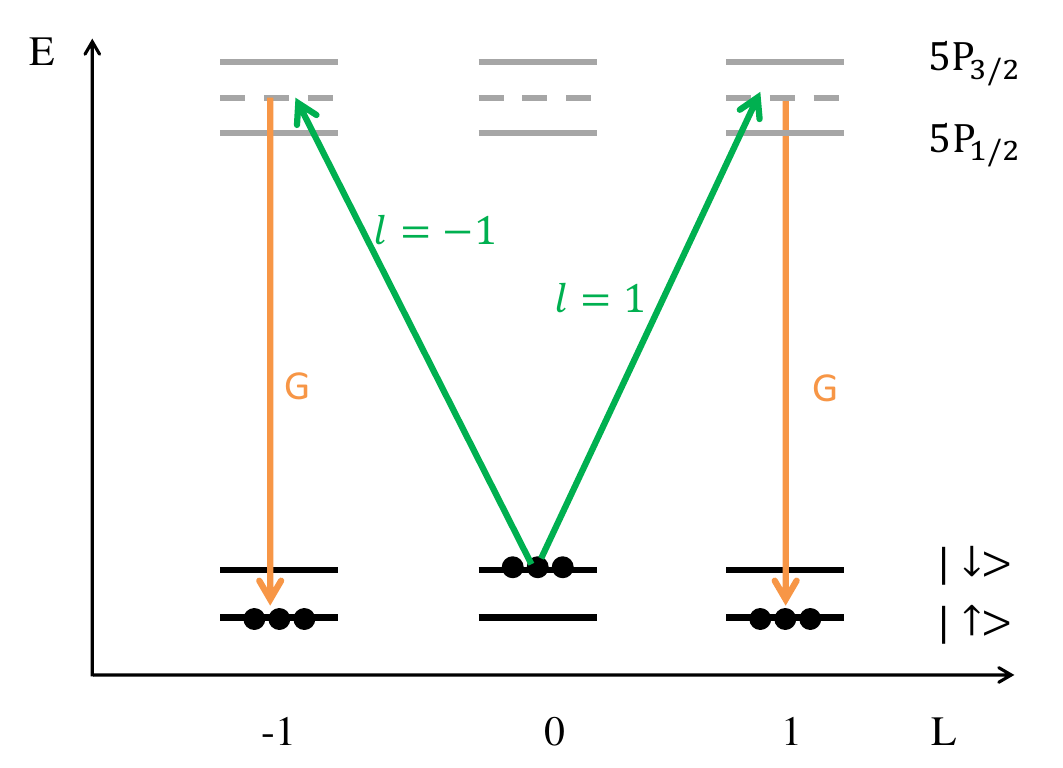}
\caption{Schematics of the coupling between atoms and Raman laser beams carrying a two-vortex superposition state. The pair of Raman laser beams are composed of a Gaussian beam (orange solid line) and a LG beam (green solid line) which carries a two-vortex superposition state with $l=\pm1$. $\ket{\downarrow}=|5S_{1/2},F=1,m_F=-1\rangle$ and $\ket{\uparrow}=|5S_{1/2},F=1,m_F=0\rangle$ are two ground internal states of atoms. The manifolds of internal states $5P_{1/2}$ and  $5P_{3/2}$ are the intermediate states of the two-photon Raman transitions. Along the horizontal axis, $L$ denotes the center-of-mass OAM of the BEC. Through the two-photon Raman transition, the atoms in the initial state $|L=0, \left\downarrow\right\rangle$ are transferred to a two-vortex superposition state composed of $|L=1, \left\uparrow\right\rangle$ and $|L=-1, \left\uparrow\right\rangle$.}
\label{sfig1}
\end{figure}
	
We use the cylindrical coordinate $\left(r,\theta, z \right)$. The Raman beams propagate along $z$ axis and the BEC locates at the plane of $z=0$. The Gaussian beam is represented as  $\vec E_G=|E_G|\vec{\epsilon}_{\sigma^-}\hat f_G\left(r\right)\cos\left(\omega_G t\right)$. The LG beam with a two-vortex superposition is denoted as $\vec E_{2\textrm{sup}}= \vec E_{+}+\alpha e^{i\phi} \vec E_{-}$ with $\vec E_{\pm}= |E_{\pm}| \vec{\epsilon}_\pi\hat f_{\pm}\left(r\right) e^{\pm i \theta} \cos\left(\omega_{\textrm{sup}} t\right)$, where $\alpha$ and $\phi$ denote the relative amplitude and relative phase between the two OAM components, respectively. $\vec{\epsilon}_{\sigma^-(\pi)}$ denotes the polarization of the Gaussian (LG) beam, and $\hat f_{G(\pm)}\left(r\right)$ is the normalized spatial distribution with $|E_{\pm}| = \frac{|E_{2\textrm{sup}}|}{\sqrt{1+\alpha^2}}$.
	
The Hamiltonian of the BEC coupled with the pair of Raman beams is written as
\begin{align}\label{eq:2supHamitaonian}
\hat H_{2\textrm{sup}} &= \sum_{L,\sigma}\varepsilon_{L,\sigma}\hat a^\dagger_{L,\sigma}\hat a_{L,\sigma} \\ \nonumber
&+\sum_{L}\sum_{\substack{\sigma_1 = \uparrow,\downarrow \\ \sigma_2 \in 5P_{1/2,3/2}}}
\left(J^G_{L,\sigma_2,\sigma_1}\hat a^\dagger_{L,\sigma_2}\hat a_{L,\sigma_1}e^{-i\omega_G t} +H.c.\right) \\ \nonumber
&+\sum_{L}\sum_{\substack{\sigma_1 = \uparrow,\downarrow \\ \sigma_2 \in 5P_{1/2,3/2}}}
\left(J^+_{L,\sigma_2,\sigma_1}\hat a^\dagger_{L,\sigma_2}\hat a_{L-1,\sigma_1}e^{-i\omega_{\textrm{sup}} t}+H.c.\right)\\ \nonumber
&+\alpha \sum_{L}\sum_{\substack{\sigma_1 = \uparrow,\downarrow \\ \sigma_2 \in 5P_{1/2,3/2}}}
\left(J^-_{L,\sigma_2,\sigma_1}\hat a^\dagger_{L,\sigma_2}\hat a_{L+1,\sigma_1}e^{-i\left(\omega_{\textrm{sup}} t-\phi\right)}+H.c.\right),
	\end{align}
where $a^{\left(\dagger\right)}_{L,\sigma}$ denotes the annihilation (creation) operator of the atom in the state of $|L,\sigma\rangle$, $\varepsilon_{L,\sigma}$ for the energy of the state, $J^G_{L,\sigma_2,\sigma_1}=\langle L,\sigma_2|\vec E_G|L,\sigma_1\rangle
=|E_G|\langle L|\hat f_G\left(r\right)|L\rangle\times\langle\sigma_2|\hat{\vec d}\Vec{\epsilon}_{\sigma^-}|\sigma_1\rangle$
defines the hopping coefficient of atoms from the ground state $|L,\sigma_1\rangle$ to the excited state $|L,\sigma_2 \rangle$ by absorbing photons of the Gaussian beam, and $\hat{\vec d}$ denotes the electric dipole moment operator. Similarly,
$J^{\pm}_{L,\sigma_2,\sigma_1}=\langle L,\sigma_2|\vec E_{\pm}|L\mp 1,\sigma_1\rangle
=|E_{\pm}|\langle L|\hat f_{\pm}\left(r\right)e^{\pm i\theta}|L\mp 1\rangle\times\langle\sigma_2|\hat{\vec d}\Vec{\epsilon}_{\pi}|\sigma_1\rangle$
is the hopping coefficient from $|L\mp 1,\sigma_1\rangle$ to $|L,\sigma_2 \rangle$ by absorbing photons of the LG beam $\vec E_{\pm}$. Since $J^{G\left(\pm\right)}_{L,\sigma_2,\sigma_1}$ is real, the hopping coefficient by emitting a photon takes the same value as that by absorbing one photon. It can also be found that $J^{+}_{L,\sigma_2,\sigma_1}=J^{-}_{L,\sigma_2,\sigma_1}$ and $\varepsilon_{L,\sigma}=\varepsilon_{-L,\sigma}$.
	
Due to the short period of the Raman pulse, the above Hamiltonian could be truncated to the Hilbert space composed of only three basis states $\{|L=0,\downarrow\rangle,|L=\pm1,\uparrow \rangle\}$ (see Fig. \ref{sfig1}), which is also valid in previous works \cite{Dowling2005PRLvortexQubit, Dowling2005PRAvortexQubit, Shi2023PRLvortexStorage}. With the standard rotating wave approximation and the adiabatic elimination of the intermediate states, the effective Hamiltonian can be written as
	\begin{equation}\label{eq:2supeffectvieHamitaonian}
		H^{\textrm{eff}}_{2\textrm{sup}}=
		\left[ \begin{array}{ccc}
			\Tilde{\varepsilon}_{0,\downarrow}               & \Omega_{+}       & \alpha\Omega_{-} e^{-i\phi} \\
			\Omega_{+}                    & \Tilde{\varepsilon}_{1,\uparrow}   &  0                      \\
			\alpha\Omega_{-} e^{i\phi}   & 0            &    \Tilde{\varepsilon}_{-1,\uparrow}
		\end{array} \right],
	\end{equation}
where $\Tilde{\varepsilon}_{0,\downarrow}=\sum_{\sigma\in 5P_{1/2,3/2}} \left(|J^{+}_{1,\sigma,\downarrow}|^2/\Delta^+_{\downarrow,\sigma}+|\alpha J^{-}_{-1,\sigma,\downarrow}|^2/\Delta^-_{\downarrow,\sigma}\right)$ and
	$\Tilde{\varepsilon}_{1,\uparrow}=\Tilde{\varepsilon}_{-1,\uparrow}=\delta+\sum_{\sigma\in 5P_{1/2,3/2}} |J^{G}_{\pm 1,\sigma,\uparrow}|^2/\Delta_{\uparrow,\sigma}$ are the renormalized energy of states $|0,\downarrow\rangle$ and $|\pm 1,\uparrow\rangle$, respectively. $\Delta^{\pm}_{\downarrow,\sigma}=\varepsilon_{0,\downarrow}-\varepsilon_{\pm 1,\sigma}+\hbar\omega_{\textrm{sup}}$, $\Delta_{\uparrow,\sigma}=\varepsilon_{1,\uparrow}-\varepsilon_{1,\sigma}+\hbar\omega_{G}$, and $\delta=\varepsilon_{\pm,\uparrow}-\varepsilon_{0,\downarrow}+\hbar\omega_{G}-\hbar\omega_{\textrm{sup}}$.
The Rabi frequency $\Omega_{\pm}= -\sum_{\sigma \in 5P_{1/2,3/2}} J^{G}_{\pm,\sigma,\uparrow}J^{\pm}_{\pm,\sigma,\downarrow} /\Delta^{\pm}_{\downarrow,\sigma}$ is given by the two-photon transition induced by the Gaussian and LG beams. Since $\Omega_{+}=\Omega_{-}$, we introduce the simplified notation $\Omega=\Omega_{\pm}$.
	Given that the density of the free-expanding BEC is dilute, the interaction between atoms is ignored in the effective Hamiltonian.
	
Initially prepared in the state $|L=0,\downarrow\rangle$, the BEC wavefunction temporally evolves as
	\begin{equation}
		|\Psi(t) \rangle =
		e^{-i \frac{\Tilde{\varepsilon}_{0,\downarrow}+ \Tilde{\varepsilon}_{1,\uparrow}}{2} t} \left( \begin{array}{c}
			\frac{\Tilde{\Omega}_{-}^2 e^{i\Omega^* t}}{\Tilde{\Omega}_{-}^2 +\left(1+\alpha^2\right)\Omega^2}
			+\frac{\Tilde{\Omega}_{+}^2  e^{-i\Omega^* t}}{\Tilde{\Omega}_{+}^2 +\left(1+\alpha^2\right)\Omega^2
			} \\
			\frac{\Omega \Tilde{\Omega}_{-}e^{i\Omega^* t}}{\Tilde{\Omega}^2_{-}+\left(1+\alpha^2\right)\Omega^2
			}
			+\frac{\Omega \Tilde{\Omega}_{+}e^{-i\Omega^* t}}{\Tilde{\Omega}^2_{+}+\left(1+\alpha^2\right)\Omega^2
			}\\
			\alpha e^{i \phi}[\frac{\Omega \Tilde{\Omega}_{-}e^{i\Omega^* t}}{\Tilde{\Omega}^2_{-}+\left(1+\alpha^2\right)\Omega^2
			}
			+\frac{\Omega \Tilde{\Omega}_{+}e^{-i\Omega^* t}}{\Tilde{\Omega}^2_{+}+\left(1+\alpha^2\right)\Omega^2
			}]
		\end{array}
		\right),
	\end{equation}
where $\Omega^{*} = \sqrt{(\frac{\Tilde{\varepsilon}_{0,\downarrow} - \Tilde{\varepsilon}_{1,\uparrow}}{2})^2 + (1+\alpha^2)\Omega^2}$ and $\Tilde{\Omega}_{\pm}=\frac{\Tilde{\varepsilon}_{0,\downarrow} - \Tilde{\varepsilon}_{1,\uparrow}}{2} \pm \Omega^{*}$. It is equivalent to write the BEC wavefunction as the following expression:
	\begin{equation}\label{eq:2supwavefunction}
		|\Psi(t) \rangle = B(t) |L=0,\downarrow\rangle + A(t) (|L=1,\uparrow\rangle + \alpha e^{i \phi} |L=-1,\uparrow\rangle)
	\end{equation}
	with $ B(t) = e^{-i \frac{\Tilde{\varepsilon}_{0,\downarrow}+ \Tilde{\varepsilon}_{1,\uparrow}}{2} t}
	[ \frac{\Tilde{\Omega}_{-}^2 e^{i\Omega^* t}}{\Tilde{\Omega}_{-}^2 +\left(1+\alpha^2\right)\Omega^2}
	+\frac{\Tilde{\Omega}_{+}^2  e^{-i\Omega^* t}}
	{\Tilde{\Omega}_{+}^2 +\left(1+\alpha^2\right)\Omega^2
	} ]$ and $ A(t) = e^{-i \frac{\Tilde{\varepsilon}_{0,\downarrow}+ \Tilde{\varepsilon}_{1,\uparrow}}{2} t}
	[\frac{\Omega \Tilde{\Omega}_{-}e^{i\Omega^* t}}{\Tilde{\Omega}^2_{-}+\left(1+\alpha^2\right)\Omega^2
	}
	+\frac{\Omega \Tilde{\Omega}_{+}e^{-i\Omega^* t}}{\Tilde{\Omega}^2_{+}+\left(1+\alpha^2\right)\Omega^2
	} ]$.
	The projection of the wavefunction on the state $\ket{\uparrow}$ is
	\begin{equation}\label{eq:2supprojection}
		\langle \uparrow | \Psi (t) \rangle = A(t) (|L=1 \rangle + \alpha e^{i \phi} |L=-1 \rangle).
	\end{equation}
Equation (\ref{eq:2supprojection}) indicates that the optical OAMs $l=\pm1$, the relative amplitude $\alpha$, and the relative phase $\phi$ in the LG beam could be well imprinted to the spin state $|\left\uparrow\right\rangle$ of the BEC.
	
\begin{figure}[htbp]
\centering
\includegraphics[width=0.8\textwidth]{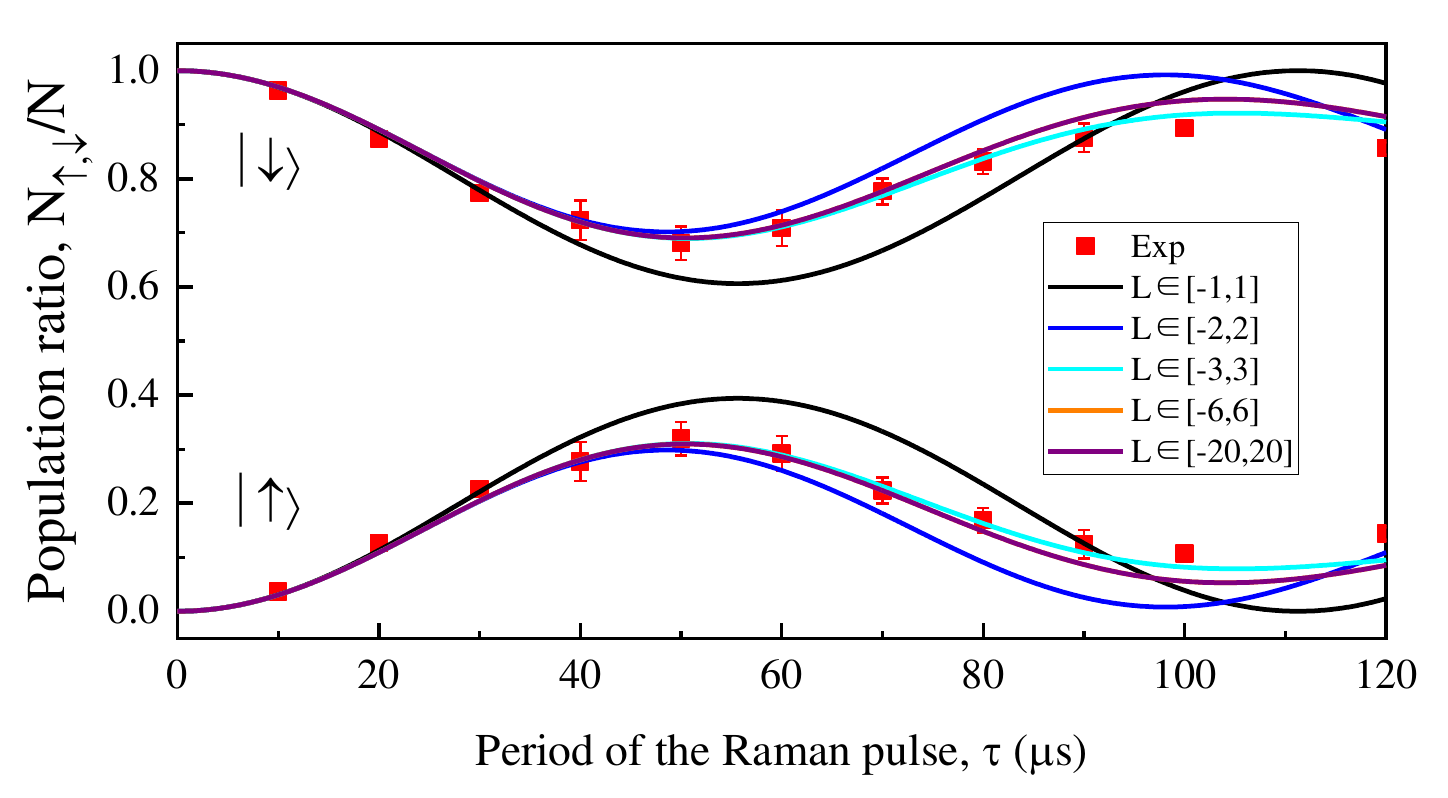}
\caption{Population ratio of the two spin states versus the period of the Raman pulse. $N_{\downarrow}$ ($N_{\uparrow}$) is the atom number of the spin state $|\left\downarrow\right\rangle$ ($|\left\uparrow\right\rangle$) and $N$ is the total atom number. The error bar is the standard deviation of five experimental measurements. The solid curves are the numerical calculations of Eq. (\ref{eq:2supHamitaonian}) using the experimental parameters for different Hilbert spaces. When the period of the Raman pulse is short, i.e., $\tau\leq50\mu \textrm{s}$, calculations for different Hilbert spaces nearly merge together, indicating that atoms mainly populate the small Hilbert space $L\in[-1,1]$. The state with larger $L$ would be populated as the Raman-pulse period increases. The calculations for $L\in[-6,6]$ and $L\in[-20,20]$ are indistinguishable when the period is equivalent to or less than hundreds of microseconds.}
\label{sfig2}
\end{figure}

\begin{figure}[htbp]
\centering
\includegraphics[width=0.8\textwidth]{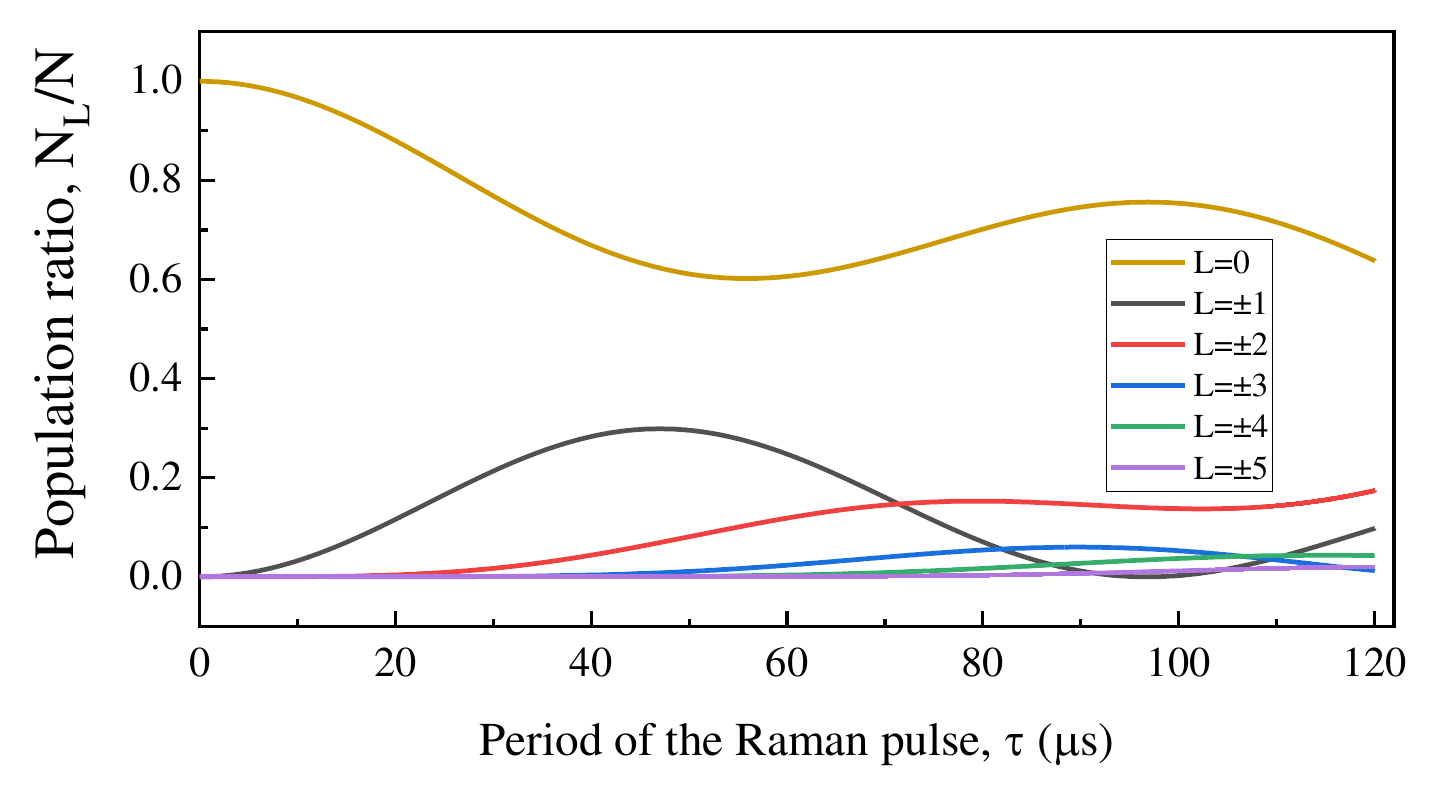}
\caption{Population ratio of the OAM states versus the period of the Raman pulse. $N_{L}$ is the atom number of the OAM state with the quantum number of $L$ and $N$ is the total atom number. The solid curves are the numerical calculations of Eq. (\ref{eq:2supHamitaonian}) using the experimental parameters in the Hilbert space $L\in[-5,5]$. As the Raman-pulse period increases, the population at $L=0$ is transferred to the state with $L=\pm1$. At $\tau \approx50 \mu \textrm{s}$, $N_{L=\pm1}$ approaches the maximum with $N_{L=\pm1}/N\approx30\%$, while $N_{L=\pm2}$ is comparably small with $N_{L=\pm2}/N\approx8\%$, and populations with larger $L$ are nearly zero. }
\label{sfig3}
\end{figure}
	
Next, we will analyze the atomic distribution among OAM states for different periods of Raman pulse. When the Raman-pulse period is short, atoms initially in the state $|L=0,\downarrow\rangle$ will first be transferred to the two states $|L=1,\left\uparrow\right\rangle$ and $|L=-1,\left\uparrow\right\rangle$, which is called the first-order two-photon Raman process and is schematically shown in Fig. \ref{sfig1}. As the Raman-pulse period increases, the atoms in states $|L=1,\left\uparrow\right\rangle$ and $|L=-1,\left\uparrow\right\rangle$ will be further transferred to the states $|L=2,\left\downarrow\right\rangle$ and $|L=-2,\left\downarrow\right\rangle$ with some possibility, which is called the second-order two-photon Raman process. Higher-order two-photon Raman processes would also occur to some extent when the Raman pulse is long. The OAM states with $L=0,\pm2,\pm4...$ exist in the spin state $|\left\downarrow\right\rangle$ and states with $L=\pm1,\pm3...$ in the spin state $|\left\uparrow\right\rangle$. In Fig. \ref{sfig2}, we measure atom numbers in the two spin states for different Raman-pulse periods. The calculations with Eq. (\ref{eq:2supHamitaonian}) are performed with different truncations of the Hilbert spaces. We find that, with the Raman-pulse period less than hundreds of microseconds, the Hilbert spaces with $L\in[-6,6]$ is large enough to cover the experimental measurements. When the Raman-pulse period is short, i.e., $\tau\leq50\mu \textrm{s}$, the calculation in the Hilbert space with $L\in[-1,1]$ has a satisfactory agreement with the measurements. We further calculate the population at different OAM states in Fig. \ref{sfig3}. As the Raman-pulse period increases, the population at $L=0$ is transferred to the state with $L=\pm1$. $N_{L}$ is the atom number of the OAM state with the quantum number of $L$ and $N$ is the total atom number. At $\tau \approx50 \mu \textrm{s}$, $N_{L=\pm1}$ approaches the maximum with $N_{L=\pm1}/N\approx30\%$, while $N_{L=\pm2}$ is comparably small with $N_{L=\pm2}/N\approx8\%$, and populations with larger $L$ are nearly zero. So, in our experiment at $\tau \approx50 \mu \textrm{s}$, about $92\%$ of atoms populate in the three states $\{|L=0,\downarrow\rangle,|L=\pm1,\uparrow \rangle\}$. In this condition, we could get a large atom number at $L=\pm1$ and the truncation of the Hilbert space to $L\in[-1,1]$ in Eq. (\ref{eq:2supeffectvieHamitaonian}) is reasonable.

\subsection{Generation of an arbitrary two-vortex superposition state}
Here, we consider that LG beam carries an arbitrary two-vortex superposition state with OAMs of $L_1$ and $L_2$, which can be written as $\vec E_{2\textrm{sup}}=  \vec E_{L_1}+\alpha e^{i\phi}\vec E_{L_2} $ with $\vec E_{L_{1(2)}}= |E_{L_{1(2)}}| \vec{\epsilon}_\pi\hat f_{L_{1(2)}}\left(r\right)e^{iL_{1(2)}\times\theta}\cos\left(\omega_{\textrm{sup}} t\right)$. Here $\hat f_{L_{1(2)}}\left(r\right)$ is the normalized spatial distribution with $|E_{L_{1(2)}}| = \frac{|E_{2\textrm{sup}}|}{\sqrt{1+ \alpha^2}}$, and $\alpha$ and $\phi$ denote the relative amplitude and relative phase between the two OAM components, respectively. The Hamiltonian expanded on the basis $\{|L,\sigma\rangle\}$ is obtained as
	\begin{align}
		\hat H_{2sup} = &\sum_{L,\sigma}\varepsilon_{L,\sigma}\hat a^\dagger_{L,\sigma}\hat a_{L,\sigma} \\ \nonumber
		+&\sum_{L}\sum_{\substack{\sigma_1 = \uparrow,\downarrow \\ \sigma_2 \in 5P_{1/2,3/2}}}
		\left(J^G_{L,\sigma_2,\sigma_1}\hat a^\dagger_{L,\sigma_2}\hat a_{L,\sigma_1}e^{-i\omega_G t} +H.c.\right) \\ \nonumber
		+&\sum_{L}\sum_{\substack{\sigma_1 = \uparrow,\downarrow \\ \sigma_2 \in 5P_{1/2,3/2}}}
		\left(J^{L_1}_{L,\sigma_2,\sigma_1}\hat a^\dagger_{L,\sigma_2}\hat a_{L-L_1,\sigma_1}e^{-i\omega_{\textrm{sup}} t}+H.c.\right)\\ \nonumber
		+\alpha &\sum_{L}\sum_{\substack{\sigma_1 = \uparrow,\downarrow \\ \sigma_2 \in 5P_{1/2,3/2}}}
		\left(J^{L_2}_{L,\sigma_2,\sigma_1}\hat a^\dagger_{L,\sigma_2}\hat a_{L-L_2,\sigma_1}e^{-i\left(\omega_{\textrm{sup}} t-\phi\right)}+H.c.\right),
	\end{align}
where the coupling coefficients are defined as $J^{L_{1\left(2\right)}}_{L,\sigma_2,\sigma_1}=\langle L,\sigma_2|\vec E_{L_{1\left(2\right)}}|L-L_{1\left(2\right)},\sigma_1\rangle
	=|E_{L_{1(2)}}|\langle L|\hat f_{L_{1\left(2\right)}}\left(r\right)e^{iL_{1(2)}\theta}|L-L_{1\left(2\right)}\rangle\times\langle\sigma_2|\hat{\vec d}\Vec{\epsilon}_{\pi}|\sigma_1\rangle$.
The effective Hamiltonian in the truncated Hilbert space spanned by $\{|0,\downarrow\rangle,|L_1,\uparrow\rangle,|L_2\uparrow\rangle\}$ reads
	\begin{equation}
		H^{\textrm{eff}}_{2\textrm{sup}}=
		\left[ \begin{array}{ccc}
			\Tilde{\varepsilon}_{0,\downarrow}                & \Omega_{L_1}       & \alpha\Omega_{L_2} e^{-i\phi} \\
			\Omega_{L_1}                    & \Tilde{\varepsilon}_{L_1,\uparrow}    &  0                      \\
			\alpha\Omega_{L_2} e^{i\phi}   & 0            &     \Tilde{\varepsilon}_{L_2,\uparrow}
		\end{array}
		\right],
	\end{equation}
where $\Tilde{\varepsilon}_{0,\downarrow}=\sum_{\sigma\in 5P_{1/2,3/2}} \left(|J^{L_1}_{L_1,\sigma,\downarrow}|^2/\Delta^{L_1}_{\downarrow,\sigma} +|\alpha J^{L_2}_{L_2,\sigma,\downarrow}|^2/\Delta^{L_2}_{\downarrow,\sigma}\right)$, and
$\Tilde{\varepsilon}_{L_{1(2)},\uparrow}=\delta+\sum_{\sigma\in 5P_{1/2,3/2}} |J^{G}_{L_{1(2)},\sigma,\uparrow}|^2/\Delta_{\uparrow,\sigma}$ are the renormalized energy of the states. Here
	$\Delta^{L_{1(2)}}_{\downarrow,\sigma}=\varepsilon_{0,\downarrow}-\varepsilon_{L_{1(2)},\sigma}+\hbar\omega_{\textrm{sup}}$
  and
	$\delta=\varepsilon_{L_{1(2)},\uparrow}-\varepsilon_{0,\downarrow}+\hbar\omega_{G}-\hbar\omega_{\textrm{sup}}$.
	The Rabi frequency $\Omega_{L_{1(2)}}= -\sum_{\sigma \in 5P_{1/2,3/2}}
	J^{G}_{L_{1(2)},\sigma,\uparrow}J^{L_{1(2)}}_{L_{1(2)},\sigma,\downarrow} /\Delta_{\downarrow,\sigma}^{L_{1(2)}}$
	refers to the two-photon transition induced by the Gaussian beam and LG beam with OAM of $L=L_{1(2)}$. Initially prepared in the state of $|L=0,\downarrow\rangle$, the BEC wavefunction temporally evolves as
	\begin{equation}
		|\Psi(t) \rangle =
		e^{-i \frac{\Tilde{\varepsilon}_{L_1,\uparrow} + \Tilde{\varepsilon}_{L_2,\uparrow}}{2} t} \left( \begin{array}{c}
			\frac{\Tilde{\Omega}_{-}^2 e^{i\Omega^* t}}{\Tilde{\Omega}_{-}^2 +\left(1+\Tilde{\alpha}^2\right)\Omega_{L_1}^2}
			+\frac{\Tilde{\Omega}_{+}^2  e^{-i\Omega^* t}}{\Tilde{\Omega}_{+}^2 +\left(1+\Tilde{\alpha}^2\right)\Omega_{L_1}^2
			} \\
			\frac{\Omega_{L_1} \Tilde{\Omega}_{-}e^{i\Omega^* t}}{\Tilde{\Omega}^2_{-}+\left(1+\Tilde{\alpha}^2\right)\Omega_{L_1}^2
			}
			+\frac{\Omega_{L_1} \Tilde{\Omega}_{+}e^{-i\Omega^* t}}{\Tilde{\Omega}^2_{+}+\left(1+\Tilde{\alpha}^2\right)\Omega_{L_1}^2
			}\\
			\alpha \frac{\Omega_{L_2}}{\Omega_{L_1}} e^{i \phi}[\frac{\Omega_{L_1} \Tilde{\Omega}_{-}e^{i\Omega^* t}}{\Tilde{\Omega}^2_{-}+\left(1+\Tilde{\alpha}^2\right)\Omega_{L_1}^2
			}
			+\frac{\Omega_{L_1} \Tilde{\Omega}_{+}e^{-i\Omega^* t}}{\Tilde{\Omega}^2_{+}+\left(1+\Tilde{\alpha}^2\right)\Omega_{L_1}^2
			}]
		\end{array}
		\right),
	\end{equation}
where $\Tilde{\alpha} = \alpha \frac{\Omega_{L_2}}{\Omega_{L_1}}$, $\Omega^{*} = \sqrt{(\frac{\Tilde{\varepsilon}_{0,\downarrow} - \Tilde{\varepsilon}_{L_{1},\uparrow}}{2})^2 + (1+\Tilde{\alpha}^2)\Omega_{L_1}^2}$ and $\Tilde{\Omega}_{\pm}=\frac{\Tilde{\varepsilon}_{0,\downarrow} - \Tilde{\varepsilon}_{L_{1},\uparrow}}{2} \pm \Omega^{*}$. It is equivalent to write the BEC wavefunction as the following expression:
	\begin{equation}
		|\Psi(t) \rangle = B(t) |L=0,\downarrow\rangle + A(t) (|L = L_1,\uparrow\rangle + \alpha \frac{\Omega_{L_2}}{\Omega_{L_1}} e^{i \phi} |L = L_2,\uparrow\rangle),
	\end{equation}
where $B(t) = e^{-i \frac{\Tilde{\varepsilon}_{L_1,\uparrow} + \Tilde{\varepsilon}_{L_2,\uparrow}}{2} t}  [ \frac{\Tilde{\Omega}_{-}^2 e^{i\Omega^* t}}{\Tilde{\Omega}_{-}^2 +\left(1+\Tilde{\alpha}^2\right)\Omega_{L_1}^2}
	+\frac{\Tilde{\Omega}_{+}^2  e^{-i\Omega^* t}}{\Tilde{\Omega}_{+}^2 +\left(1+\Tilde{\alpha}^2\right)\Omega_{L_1}^2
	} ] $ and $ A(t) = e^{-i \frac{\Tilde{\varepsilon}_{L_1,\uparrow} + \Tilde{\varepsilon}_{L_2,\uparrow}}{2} t}  [ \frac{\Omega_{L_1}\Tilde{\Omega}_{-}e^{i\Omega^* t}}{\Tilde{\Omega}^2_{-}+\left(1+\Tilde{\alpha}^2\right)\Omega_{L_1}^2
	}
	+\frac{\Omega_{L_1}\Tilde{\Omega}_{+}e^{-i\Omega^* t}}{\Tilde{\Omega}^2_{+}+\left(1+\Tilde{\alpha}^2\right)\Omega_{L_1}^2
	} ] $.
	The projection of the wavefunction on the state $\ket{\uparrow}$ is
	\begin{equation}\label{eq:2supprojectionV2}
		\langle \uparrow | \Psi (t) \rangle = A(t) (|L = L_1 \rangle + \alpha \frac{\Omega_{L_2}}{\Omega_{L_1}} e^{i \phi} |L = L_2 \rangle).
	\end{equation}
Equation (\ref{eq:2supprojectionV2}) indicates that the optical OAMs $l=L_{1,2}$ and the relative amplitude $\phi$ in the LG beam could be well imprinted to the spin state $|\left\uparrow\right\rangle$ of the BEC. The relative amplitude $\alpha$ with a correction of $\frac{\Omega_{L_2}}{\Omega_{L_1}}$ could also be transferred to the BEC. Here, $\frac{\Omega_{L_2}}{\Omega_{L_1}}$ mainly depends on the spatial distribution of the two OAM components. If $L_{1}=1$ and $L_{2}=-1$, $\Omega_{L_1}=\Omega_{L_2}$, and then Eq. (\ref{eq:2supprojectionV2}) will become Eq. (\ref{eq:2supprojection}).

\subsection{Generation of a three-vortex superposition state}
We will consider that there is a three-vortex superposition state $l=0, \pm 1$ in the LG beam. One example of this superposition state is $\vec E_{3\textrm{sup}}=\vec E_+ +e^{i\pi}\vec E_-+\alpha e^{i\phi}\vec E_{G'}$, where $\alpha$ and $\phi$ are the relative amplitude and relative phase of the additional Gaussian component, respectively. The additional Gaussian component is given as $\vec E_{G'}= |E_{G'}| \vec \epsilon_{\pi} \hat f_{G'}\left(r\right)\cos\left(\omega_{\textrm{sup}}t\right)$ and $|E_{\pm}| = |E_{G'}| = \frac{|E_{3\textrm{sup}}|}{\sqrt{2+\alpha^2}}$.
The Hamiltonian of the BEC coupled with the pair of Raman beams could be written as
	\begin{align}\label{eq:3supHamitaonian}
		\hat H_{3\textrm{sup}} = &\sum_{L,\sigma}\varepsilon_{L,\sigma}\hat a^\dagger_{L,\sigma}\hat a_{L,\sigma} \\ \nonumber
		+&\sum_{L}\sum_{\substack{\sigma_1 = \uparrow,\downarrow \\ \sigma_2 \in  5P_{1/2,3/2}}}
		\left(J^G_{L,\sigma_2,\sigma_1}\hat a^\dagger_{L,\sigma_2}\hat a_{L,\sigma_1}e^{-i\omega_G t} +H.c.\right)\\ \nonumber
		+&\sum_{L}\sum_{\substack{\sigma_1 = \uparrow,\downarrow \\ \sigma_2 \in  5P_{1/2,3/2}}}
		\left(J^+_{L,\sigma_2,\sigma_1}\hat a^\dagger_{L,\sigma_2}\hat a_{L-1,\sigma_1}e^{-i\omega_{\textrm{sup}} t}+H.c.\right)\\ \nonumber
		-&\sum_{L}\sum_{\substack{\sigma_1 = \uparrow,\downarrow \\ \sigma_2 \in  5P_{1/2,3/2}}}
		\left(J^-_{L,\sigma_2,\sigma_1}\hat a^\dagger_{L,\sigma_2}\hat a_{L+1,\sigma_1}e^{-i\omega_{\textrm{sup}} t}+H.c.\right)\\ \nonumber
		+\alpha &\sum_{L}\sum_{\substack{\sigma_1 = \uparrow,\downarrow \\ \sigma_2 \in  5P_{1/2,3/2}}}
		\left(J^{G'}_{L,\sigma_2,\sigma_1}\hat a^\dagger_{L,\sigma_2}\hat a_{L,\sigma_1}e^{-i\left(\omega_{\textrm{sup}} t-\phi\right)}+H.c.\right).
	\end{align}
Compared to $\hat H_{2\textrm{sup}}$ in Eq. (\ref{eq:2supHamitaonian}), Eq. (\ref{eq:3supHamitaonian}) includes an additional transition channel with the hopping coefficient $J^{G'}_{L,\sigma_2,\sigma_1}=\langle L,\sigma_2|\vec E_{G'}|L,\sigma_1\rangle=|E_{G'}|\langle L|\hat f_{G'}\left(r\right)|L\rangle\times\langle\sigma_2|\hat{\vec d}\Vec{\epsilon}_{\pi}| \sigma_1\rangle$, which is induced by the additional Gaussian component in the LG beam.
\begin{figure}[htbp]
\centering
\includegraphics[width=0.8\textwidth]{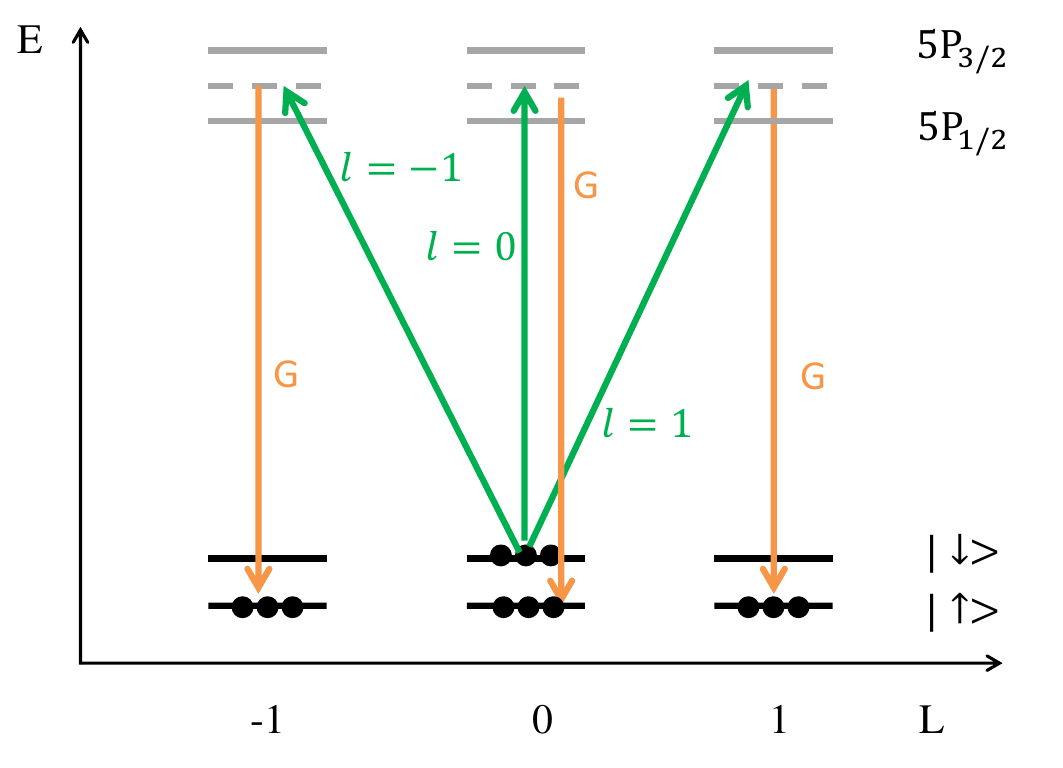}
\caption{Schematics of the coupling between atoms and the Raman laser beams carrying a three-vortex superposition state. The pair of Raman laser beams are composed of a Gaussian beam (orange solid line) and a LG beam (green solid line) which carries a three-vortex superposition state with $l=0, \pm1$. $\ket{\downarrow}=|5S_{1/2},F=1,m_F=-1\rangle$ and $\ket{\uparrow}=|5S_{1/2},F=1,m_F=0\rangle$ are two ground internal states of atoms. The manifolds of internal states $5P_{1/2}$ and  $5P_{3/2}$ are the intermediate states of the two-photon Raman transitions. Along the horizontal axis, $L$ denotes the center-of-mass OAM of the BEC. Through the two-photon Raman transition, the atoms initially in the state $|L=0, \left\downarrow\right\rangle$ is transferred to a three-vortex superposition state composed of $|L=1, \left\uparrow\right\rangle$, $|L=0, \left\uparrow\right\rangle$, and $|L=-1,\left\uparrow\right\rangle$.}
\label{sfig4}
\end{figure}

Under the first-order two-photon Raman process with a short period of Raman pulse, the Hilbert space is truncated to the basis of $\{|L=0,\downarrow\rangle, |L=0,\uparrow\rangle, |L=\pm 1,\uparrow\rangle\}$. As shown in Fig. \ref{sfig4}, atoms initially in the state $|L=0,\downarrow\rangle$ will be transferred to three states $|L=1, \left\uparrow\right\rangle$, $|L=0,\left\uparrow\right\rangle$, and $|L=-1,\left\uparrow\right\rangle$. The effective Hamiltonian is written as
	\begin{equation}
		H^{\textrm{eff}}_{3\textrm{sup}}=
		\left[ \begin{array}{cccc}
			\Tilde{\varepsilon}_{0,\downarrow}        &\Omega_{+}         &\alpha \Omega_{0} e^{-i \phi}       & -\Omega_{-} \\
			\Omega_{+}     & \Tilde{\varepsilon}_{1,\uparrow}        & 0     & 0  \\
			\alpha \Omega_{0} e^{i \phi}    & 0        & \Tilde{\varepsilon}_{0,\uparrow}     & 0   \\
			- \Omega_{-}        & 0        & 0     & \Tilde{\varepsilon}_{-1,\uparrow}
		\end{array}
		\right],
	\end{equation}
where $\Tilde{\varepsilon}_{0,\downarrow}=\sum_{\sigma\in 5P_{1/2,3/2}} \left(|J^{+}_{1,\sigma,\downarrow}|^2/\Delta^+_{\downarrow,\sigma} +|J^{-}_{-1,\sigma,\downarrow}|^2/\Delta^-_{\downarrow,\sigma}+|\alpha J^{G'}_{0,\sigma,\downarrow}|^2/\Delta^{G'}_{\downarrow,\sigma}\right)$ and
$\Tilde{\varepsilon}_{0,\uparrow}=\delta_0+\sum_{\sigma\in 5P_{1/2,3/2}} |J^{G'}_{0,\sigma,\uparrow}|^2/\Delta_{\uparrow,\sigma}$. Here,
	$\Delta^{G'}_{\downarrow,\sigma}=\varepsilon_{0,\downarrow}-\varepsilon_{0,\sigma}+\hbar\omega_{\textrm{sup}}$,
	$\Delta_{\uparrow,\sigma}=\varepsilon_{0,\uparrow}-\varepsilon_{0,\sigma}+\hbar\omega_{G}$ and
	$\delta_0=\varepsilon_{0,\uparrow}-\varepsilon_{0,\downarrow}+\hbar\omega_G-\hbar\omega_{\textrm{sup}}$.
The Rabi frequency of the additional channel induced by the Gaussian beam and the Gaussian component in the LG beam can be written as
	$\Omega_{0}=-\sum_{\sigma \in 5P_{1/2,3/2}}J^{G}_{0,\sigma,\uparrow} J^{G'}_{0,\sigma,\downarrow}/\Delta^G_{\uparrow,\sigma}$.
	
Initially prepared in the state $|L=0,\downarrow\rangle$, the BEC wavefunction temporally evolves as
	\begin{equation}
		|\Psi(t) \rangle =
		\left( \begin{array}{c}
			B(t)  \\
			A(t)  \\
			A(t) \alpha \frac{ \Omega_{0} }{ \Omega } e^{i \phi}  \\
			-A(t)
		\end{array}
		\right),
	\end{equation}
where $A(t) = \frac{1}{c_3^2} \frac{- \delta' - \Delta'}{2 \Omega} e^{-i \frac{\Omega_0}{2} (\delta' - \Delta') t } + \frac{1}{c_4^2} \frac{- \delta' + \Delta'}{2 \Omega} e^{-i \frac{\Omega_0}{2} (\delta' + \Delta') t } $, $B(t) = \frac{1}{c_3^2} | \frac{- \delta' - \Delta'}{2 \Omega} |^2 e^{-i \frac{\Omega_0}{2} (\delta' - \Delta') t } +  \frac{1}{c_4^2} | \frac{- \delta' + \Delta'}{2 \Omega} |^2 e^{-i \frac{\Omega_0}{2} (\delta' + \Delta') t } $,
	$\delta' = \delta - \varepsilon_{0,\downarrow}$, $\Delta' = \sqrt{\delta'^2 + 4 \Omega_0^2 \alpha^2 + 8 \Omega^2}$. $c_3 = \sqrt{(\frac{-\delta' -\Delta' }{2 \Omega})^2 + 2 +(\frac{\alpha \Omega_0}{\Omega})^2}$, and $c_4 = \sqrt{(\frac{-\delta' + \Delta' }{2 \Omega})^2 + 2 +(\frac{\alpha \Omega_0}{\Omega})^2}$.
	It is equivalent to write the BEC wavefunction as the following expression:
	\begin{equation} \label{eq:3supwavefunction}
		\begin{split}
			|\Psi(t) \rangle &= B(t) |L=0,\downarrow\rangle +
			A(t) (|L=1,\uparrow\rangle + \alpha \frac{ \Omega_{0} }{ \Omega } e^{i \phi} |L=0,\uparrow\rangle +e^{i\pi} |L=-1,\uparrow\rangle  ).
		\end{split}
	\end{equation}
	The projection of the wavefunction on the state $\ket{\uparrow}$ is
	\begin{equation}\label{eq:3supprojection}
		\langle \uparrow | \Psi (t) \rangle = A(t) (|L=1 \rangle + \alpha \frac{ \Omega_{0} }{ \Omega } e^{i \phi}  |L=0 \rangle + e^{i\pi} |L=-1 \rangle  ).
	\end{equation}
Equation (\ref{eq:3supprojection}) indicates that the optical OAMs $l=0,\pm1$ and the relative amplitude $\phi$ in the LG beam could be well imprinted to the spin state $|\left\uparrow\right\rangle$ of the BEC. The relative amplitude $\alpha$ with a correction of $\frac{\Omega_{0}}{\Omega}$ could also be transferred to the BEC. Here, $\frac{\Omega_{0}}{\Omega}$ mainly depends on the different spatial distributions of the Gaussian ($l=0$) and vortex ($l=\pm1$) components in the LG beam.

\section{Representation of a three-state superposition using two Bloch spheres} \label{section:VI}
Without loss of generality, we write a three-state superposition as
\begin{equation}
	|\psi\rangle= |1\rangle + a_2 e^{i \phi_2}|2\rangle + a_3 e^{i \phi_3}|3\rangle 
\end{equation}
where $a_{2,3}$ are relative amplitudes and $\phi_{2,3}$ are relative phases. Normalizing the above wavefunction, we have
\begin{eqnarray} \label{eq:2Blochsphere} 
	\nonumber |\psi\rangle &=& |1\rangle + a_2 e^{i \phi_2}|2\rangle + a_3 e^{i \phi_3}|3\rangle \\
	\nonumber &=& \sqrt{1+a_2^2+a_3^2}\left[ \frac{\sqrt{1+a_2^2}}{\sqrt{1+a_2^2+a_3^2}} \left( \frac{1}{\sqrt{1+a_2^2}} |1\rangle +\frac{a_2}{\sqrt{1+a_2^2}} e^{i \phi_2}|2\rangle \right) 
	+ \frac{a_3}{\sqrt{1+a_2^2+a_3^2}} e^{i \phi_3}|3\rangle \right] \\
	&=& \sqrt{1+a_2^2+a_3^2}\left[ \sin \frac{\theta_{3}}{2}\Bigg( \sin\frac{\theta_{2}}{2}  |1 \rangle + \cos\frac{\theta_{2}}{2}e^{i \phi_2}|2\rangle  \Bigg)+\cos\frac{\theta_{3}}{2}e^{i \phi_3}|3\rangle \right].
\end{eqnarray}
Here $\theta_{2,3} \in [0,\pi)$ and $\phi_{2,3} \in [0,2\pi)$. Equation \eqref{eq:2Blochsphere} give the relation between the three-state superposition and the two Bloch spheres. $\theta_{2}$ and $\phi_{2}$ are the latitude and longitude of the Bloch sphere 1, and $\theta_{3}$ and $\phi_{3}$ are the latitude and longitude of the Bloch sphere 2, respectively. The latitudes $\theta_{2,3}$ are obtained as 
\begin{eqnarray}
	\sin\frac{\theta_{2}}{2}&=&\frac{1}{\sqrt{1+a_2^2}}, \\
	\sin\frac{\theta_{3}}{2}&=&\frac{\sqrt{1+a_2^2}}{\sqrt{1+a_2^2+a_3^2}}.
\end{eqnarray}
This means that $\theta_{2,3}$ and $\phi_{2,3}$ are uniquely determined for a known three-state superposition. The latitude $\theta_{2}$ only depends on the relative amplitude $a_{2}$, and $\theta_{3}$ depends on $a_{2}$ and $a_{3}$. Using this method, we represent a three-vortex superposition state with two Bloch spheres in Fig. 5 of the main text and in Fig. \ref{figS3}.


%

\end{widetext}
	
\end{document}